%% file: ms.tex
\documentclass[12pt,preprint]{aastex}
\input psfig.sty

\def\h1{\mbox {\rm HI}}
\def\x{\mbox {\rm X-ray~}}
\def\X{\mbox {\rm X-Ray~}}
\def\fig{{Figure}}

\def\nh{\mbox {${\rm N}_{\rm H}$}~}
\def\nhg{\mbox {${\rm N}_{\rm H,Gal}$}~}
\def\o16{\mbox {$^{16}O$~}}

\def\deg{\mbox {$^{\circ}$}~}
\def\msun{\mbox {${\rm ~M_\odot}$}}
\def\zsun{\mbox {${\rm ~Z_{\odot}}$}}

\def\osun{\mbox {${\rm ~(O/H)_{\odot}}$}}
\def\lsun{\mbox {${~\rm L_\odot}$}}
\def\msunyr{\mbox {$~{\rm M_\odot}$~yr$^{-1}$}}
\def\msunyrk2{\mbox {$~{\rm M_\odot}$~yr$^{-1}$~kpc$^{-2}$}}
\def\msunpc2{\mbox {${\rm ~M_\odot ~pc}^{-2}$}}
\def\angs{\mbox {~\AA}}
\def\Ha{\mbox {H$\alpha$~}}

\def\lines{\mbox {~$\lambda\lambda$~}}
\def\n{NGC~}
\def\asec{\ifmmode {'' }\else $''~$\fi}  
\def\amin{\ifmmode {' }\else $'~$\fi}    
\def\sles{\lower2pt\hbox{$\buildrel {\scriptstyle <}
   \over {\scriptstyle\sim}$}} 
\def\sgreat{\lower2pt\hbox{$\buildrel {\scriptstyle >}
    \over {\scriptstyle\sim}$}} 
\def\col{\mbox {\rm ~cm$^{-2}$} }
\def\norm{\mbox {\rm ~cm$^{-5}$} }
\def\kms{\mbox {~km~s$^{-1}$} }
\def\ergsec{~ergs~s$^{-1}$~}

\def\flux{~erg~s$^{-1}$~cm$^{-2}$}
\def\nden{\mbox {~cm$^{-3}$} }
\def\cm5{~cm$^{-5}$}

\def\et{{\rm et\thinspace al.}\ }   

\def\apj{ApJ}
\def\apjs{ApJS}

\def\aj{AJ}
\def\mn{MNRAS}

\def\aa{A\&A}

\begin{document}

\title{The Metal Content of Dwarf Starburst Winds:  Results
from Chandra Observations of NGC 1569\altaffilmark{1}}

\shorttitle{Metal Content of Starburst Winds}
\shortauthors{Martin, Kobulnicky, Heckman}

\author{Crystal L. Martin\altaffilmark{2,3}}
\affil{Caltech}
\affil{Mail stop 105-24, Pasadena CA 91125}
\email{clm@astro.caltech.edu}

\author{Henry A. Kobulnicky}
\affil{Astronomy Department, University of Wisconsin - Madison}
\affil{475 N. Charter St., Madison, WI 53706}
\email{chip@astro.wisc.edu}

\author{Timothy M. Heckman}
\affil{The Johns Hopkins University}
\affil{Department of Physics \& Astronomy}
\affil{Baltimore, MD, 21218}

\altaffiltext{1}{Based on observations obtained with the Chandra X-ray 
Observatory}

\altaffiltext{2}{Also Physics Department, University of California Santa 
Barbara}

\altaffiltext{3}{Visiting astronomer National Optical Astronomy
Observatory (NOAO). NOAO is operated by the Association of 
Universities for Research in Astronomy (AURA), Inc. under 
cooperative agreement with the National Science Foundation.}


\begin{abstract}
We present deep, Chandra spectral imaging of the dwarf starburst
galaxy \n1569.  The unprecedented spatial resolution allows us to
spatially identify the components of the integrated \x spectrum.  
Fitted spectral  models require an intrinsic absorption component and 
higher metal abundances than previous studies indicated.  Our results provide 
the first direct evidence for metal-enriched winds from dwarf starburst
galaxies.

We identify 14 \x point sources in \n1569. Most have properties 
consistent with those of high mass \x binaries, but one is a steep-spectrum 
radio source which is probably a supernova remnant.  
The \x luminosity of \n1569 is dominated by diffuse, thermal emission
from the disk ($0.7$~keV) and bipolar halo ($0.3$~keV).  Photoelectric 
absorption from the inclined HI disk hardens the \x spectrum on the 
northern side of the disk relative to the southern side.  Requiring the 
fitted absorption column to match the HI column measured at 21-cm 
implies that the metallicity of the HI disk is significantly less than solar but
greater than 0.1\zsun.  Hence, much of the HI is enriched to levels
comparable to the metallicity of the HII regions (O/H = 0.2\osun).  The
\x color variations in the halo are inconsistent with a free-streaming wind 
and  probably reveal the location of shocks created by the interaction of the 
wind with a gaseous halo. The \x spectrum of the diffuse gas presents strong 
emission lines from alpha-process elements. Fitted models require alpha
element abundances greater than 0.25$~Z_{\alpha, \odot}$ and ratios of
alpha elements to iron 2 to 4 times higher than the solar ratio. The best
fit to the spectrum is obtained with solar mass fractions for the alpha
elements, $1.0~Z_{\alpha, \odot}$; but a degeneracy between the metallicity 
and the spectral normalization prevents us from deriving an upper limit on the
 wind metallicity from the \x spectrum alone.  We argue, however, that  
abundances larger than $2.0~Z_{\alpha, \odot}$ pose awkward implications for
the dynamical evolution of the wind based on our knowledge of the starburst
properties.  For consistency with our best fitting abundances, the
mass of interstellar gas entrained in the wind must be about nine times
the mass of stellar ejecta in the wind.  Most of the oxygen carried by the 
wind comes from the stellar ejecta, rather than entrained interstellar gas.
The estimated mass of oxygen in the hot wind, 34,000\msun, is similar to the
oxygen yield of the current starburst. {\it Apparently the wind carries 
nearly all 
the metals ejected by the starburst.} These metals appear destined to contribute
to the enrichment of the intergalactic medium. Much of the nucleosynthesis
in \n1569 must have occurred during less violent periods of star formation, 
however, because our measurements imply the neutral gas disk holds at least 
five times more oxygen than wind.


%
%
%

\end{abstract}

\subjectheadings{galaxies: formation --- galaxies: evolution ---
galaxies: fundamental parameters --- galaxies: abundances }

\section{Introduction}

%
 
 Galactic winds have been advocated as a mechanism which disperses  metals 
throughout galactic disks and the intergalactic medium (IGM).  This paper
 presents new Chandra observations of the dwarf galaxy \n1569 which
 demonstrate that much of the metal-rich ejecta produced in the
 current starburst must reside in the hot \x wind.   This result
 is of particular interest because the hot gas is not gravitationally
 bound to this intermediate mass dwarf galaxy (e.g., Heckman \et 1995;
 Martin 1999).

\subsection{\n1569}

The nearby dwarf galaxy \n1569 entered a starburst phase 10 to 20~Myr
ago (Israel \& deBruyn 1988; Gonzalez-Delgado \et 1997; Hunter \et
2000).  An encounter with a low-mass cloud of HI ($7 \times 10^6$\msun\
and 5~kpc away) in the IC~342 Group may have triggered the
current activity (Stil \& Israel 1998).  The nearest galaxy with
similar radial velocity, UGCA~92, lies at a projected distance of
60~kpc (Karachentsev, Tikhonov, \& Sazonova 1994).  The young stellar
population spreads across the disk, which is inclined 60\deg\ to our
sightline (Israel 1998), but is concentrated in the center where
several super star clusters are found (O'Connell \et 1994; Hunter \et
2000).  The two brightest clusters alone have probably produced several
thousand supernovae, and the total starburst  $\sim 30,000$.  The
energy imparted to the interstellar medium by these supernova
explosions drives a bipolar outflow seen extending to either side of
the disk in \Ha emission (Heckman \et 1995; Martin 1998).

The coronal gas in the outflow was imaged in soft X-rays with ROSAT 
(Heckman \et 1995), and ASCA spectra confirmed the emission mechanism
was thermal (Della Ceca \et 1996).  The fitted ASCA temperature, 0.6~keV
for MEKAL models or 0.8~keV for Raymond-Smith models (Della Ceca \et 1996), 
was shown to be several times higher than the escape temperature from
the dark matter halo (Martin 1999). Since the cooling time of the hot gas, 
about 100 Myr, is longer than the dynamical age of the outflow, about 10 Myr,
the \x emitting gas appears likely to escape (e.g., Heckman \et 1995).  In 
order for the wind to stall, it must either encounter a very large mass of
halo gas or undergo extensive mixing with cooler gas which would accelerate the
radiative losses.

If this wind carries the bulk of the metals expelled by supernovae 
in the starburst population, it could solve a longstanding puzzle.
The HII regions in \n1569 show no local enhancements of O, N, or He as would be
expected if the material ejected from massive stars was mixed directly into
the HII regions (Kobulnicky \& Skillman 1997).  This lack of enrichment
is most conspicuous in low metallicity galaxies (i.e., large $dZ/Z$) where
it is  common (Tenorio-Tagle 1996; Martin 1996; Kobulnicky \& Skillman 1997).  
The galaxy \n1569 is particularly valuable as a case-study because of its 
proximity (d = 2.2~Mpc, Israel 1988),
low escape velocity (80-110\kms, Martin 1999), and low metallicity
(0.20 \osun, Kobulnicky \& Skillman 1997; Martin 1997).  It is caught
in a starburst phase, but otherwise is a typical dwarf with dynamical
mass  $M_{dyn} \sim 2 \times  10^8$\msun\ (e.g., Martin 1998) and 
$L_{Bol} \approx 1.7 \times 10^9$\lsun\ (Israel 1988).


\subsection{Abundance Determinations from X-Ray Spectra}

In principle, abundance determinations for a collisionally-ionized
plasma are easier than in the nebular case where photoionization must
be included in the model.  Under coronal conditions, the ionization structure 
is completely determined by the electron temperature which is fixed by external 
processes, the hydrodynamic shocks in the case of galactic winds.
Metal abundances derived from thermal \x spectra have a controversial
history, however, due to uncertainties in the atomic physics, different
definitions of the solar abundance, and a number of degeneracies inherent
to multi-component spectral models.

Reports of very sub-solar abundances in elliptical galaxies,
for example, were met with skepticism, and have been traced in part to
uncertainties in the Fe-L shell atomic physics (Liedahl \et 1995).
A degeneracy between high metallicity, two-temperature models and low
metallicity, single-temperature models has also produced reports of
surprisingly low metallicities (Buote \& Canizares 1994).  
In spectra of galaxy clusters, relative abundances have proven difficult to 
constrain owing to blends between Fe-L lines and lines from alpha-process 
elements (Mushotzsky \et 1996).  Hence,  the metallicities tend to be pushed 
lower than their true value if the temperature range is not fully represented.
Weaver \et (2000), see also Dahlem \et 1998, 
 have argued that supersolar $[\alpha/Fe]$ ratios in \n253 
and M82 are ramifications of underestimating the absorbing column.  Intrinsic
absorption omitted from the spectral model will increases the apparent 
strength of Mg and Si lines relative to Fe-L lines.  Unless lower energy 
lines from O are detected, a statistically acceptable fit can be obtained with
an anomalously high alpha element to Fe abundance ratio.
In spite of these common traps, we will argue
that analysis of the Chandra data yield (1) a unique alpha to iron abundance
ratio and (2) significantly constrain the absolute abundance of the outflow.

We adopt the MEKAL models implemented in XSPEC~10 (Arnaud 1996) which include 
the Fe L-shell calculations of Liedahl \et (1995).  They have been shown
to provide accurate temperatures for elliptical galaxies with $kT = 0.7$ to 
1.0~keV (Buote \& Fabian 1997) and galaxy clusters (Mushotzsky \et 1996).
Since the line emissivity increases relative to the continuum as the
plasma temperature decreases from a few keV to a few tenths of a keV
(e.g., Raymond \& Smith 1977), it is somewhat easier to compare the 
alpha-element line strengths to the Fe-L lines in spectra of starburst galaxies
which have cooler spectra.
At temperatures $\sim 0.5$~keV, we find that relatively small, $\ge\ 10\% $, 
changes in metal abundance produce significant variations in the count 
distribution at the spectral resolution of the back illuminated 
Chandra CCD's ($\sim 100$~eV).

In \S 2, we describe the Chandra observations and optical and radio 
observations.  The latter define the amount and location of cooler gas as 
well as the geometry of the system -- information that proves essential for
breaking the spectral degeneracies. In \S 3, we use the spectral imaging 
capabilities of Chandra to isolate \x emission and absorption components.
The integrated spectrum proves to be so complex, that this approach is
the only way to build a physical model that breaks the various degeneracies
summarized above. In \S 4, we derive the chemical properties of the \x
emitting gas which are then improved upon when the dynamics of the outflow
are considered jointly  in \S 5.  The main conclusions are summarized
in \S 6 and have important implications for the chemical
evolution of dwarf galaxies, the dynamics of winds, and the pollution
of the IGM.


\section{Data Acquisition and Reduction}
\subsection{Chandra Observations} \label{sec:xdata}

The {\em Chandra} X-ray Observatory observed \n1569 with the AXAF CCD
Imaging Spectrometer (ACIS) on 2000 April 11 for 96.8 ksec.  The galaxy
was placed on the back-illuminated chip ACIS-S3 (also known as chip~7).  
These data as well as the fields on the ACIS-I and ACIS-S5 chips are
available in the archive under sequence number 600085.

The data were processed by the Chandra X-Ray Center (CXC) using version 
R4CU5UPD13.1 of their software. We checked the positional accuracy of
the chip~7 Chandra sources using our R-band image (Section 2.2) and 20 cm 
radio map (Section 2.3). For the 16 X-ray sources  with optical detections and 
3 X-ray sources with radio counterparts, the maximum positional discrepancy
was found to be 0.5\arcsec.  Reprocessing with {CIAO}~version 2.1.3
was required to make the gainfile consistent with the revisions of 2001 August 
to the ACIS-S3 FEFs (for focal plane temperature of -120 C).  The shifted 
energy scale and decreased spectral 
resolution, significantly improved the fit statistics of our best models.
Periods of high background were defined by time intervals with count rates 
differing by more than a factor of 1.2 from the mean quiescent background 
rate on chip~7, i.e., 1.55 cnts/s. Eliminating these flares improved the
signal-to-noise ratio of the data and reduced the net exposure time to 
84.89 ksec.



\subsubsection{Spectral Extraction}

Spectra were extracted using pulse invariant data values to account for
gain variations between nodes.  For each source region, we weighted the 
appropriate CXC spectral response files by the distribution of their areas
within the aperture.\footnote{We used the calcrmf/calcarf software package
    contributed by Jonathon McDowell which is
    available from the CXC website -- 
    http://asc.harvard.edu/cgi-gen/cont-soft/soft-list.cgi.}
The maximum difference between our area-weighted response files and 
single response files extracted at the flux-weighted centroid of the
aperture was a few percent. Weighting the response files by the distribution
of counts within the aperture produced results indistinguishable from the
area-weighted responses.

A background events file was constructed from deep ACIS observations of blank 
fields using software provided by M. Markevitch through the CXC.  After
normalizing to the net exposure time of the \n1569 observations,  the
source free regions of our data exhibit a count rate $\sim 4\%$ higher
than the same regions in the blank-field background map.  This discrepancy is 
consistent with the known slow, temporal decrease in the ACIS-S3 background 
rate described by Markevitch.  After correcting the blank-field background 
rates by this factor of 1.04, we estimate the mean background rate is good
to $\pm 1\%$ across the entire field.


\subsubsection{Image Extraction} \label{sec:bands}

Unbinned images were extracted from the time-filtered events file in four 
carefully chosen bands: Soft (S) 0.3 -- 0.7~keV, Medium (M) 0.7 -- 1.1~keV,
Hard (H) 1.1 -- 6~keV, and total 0.3 -- 6~keV.  
Emission in the soft band is highly attenuated because the Galactic 
foreground column, $N_{H} = 2.1 \times 10^{21}\col$ (Burton 1985), presents 
optical depth unity at 0.7~keV and $\tau = 6.5$ at 0.3~keV.  The galaxy was 
not detected at energies above 6~keV, the energy adopted for the Hard band
cutoff. Background images were extracted from the background events file, 
described above, in these bands.

Each of the four X-ray images was smoothed using the adaptive
smoothing algorithm of H. Ebeling \& V. Rangarajan as implemented in
CSMOOTH (CIAO 2.0).   The smoothing scales are automatically adjusted
to achieve a minimum S/N ratio of 2 and a maximum S/N ratio of 3 per
pixel.  Strong point sources are effectively unchanged by the smoothing
process, while the contrast of weak, diffuse emission is enhanced.
Each of the four background images was smoothed in the same manner and
subtracted from the source images.

Since adaptive smoothing does not preserve photon statistics in a
straightforward manner, we used the adaptively smoothed images only to
produce images for presentation and to define apertures for the
extraction of spectra.  All quantitative analyses was performed by extracting 
photons from the events file or unsmoothed images and estimating their
significance using Poisson statistics.


\subsubsection{Source Identification} \label{sec:detect}

The CIAO 2.0  point source detection routine {\it celldetect} was run
on the (unsmoothed) broadband 0.3-6 keV image to identify point-like sources.
Visual examination of the resulting detections showed that some sources
with a significance level as low as 1.9 $\sigma$ in the central portion
of the chip corresponded to R-band sources (i.e., stars or
background AGN).  After several iterations of \x detection, image 
inspection, and optical counterpart searching, we adopted the following
approach.  In the central 1\arcmin\ of the S3 chip where the Chandra
PSF is best, we set a minimum detection threshold of 1.9 $\sigma$.  For
a 100 ks exposure, this threshold should lead to less than 1 false
detection.\footnote{ See the Chandra {\it Detect 1.0}  users guide
Section 4.6  for approximate detection thresholds as a function of
false detection rate.}  For the region between 1 and 2.5 \arcmin\ from
the center of the chip, a detection threshold of 2.1 was used to
achieve the same probable false detection rate.  For the section of the
chip beyond 2.5\arcmin\ from the center, we used a detection threshold
of 3.2$\sigma$.  The point spread function degrades significantly
beyond 3.5\arcmin, so we did not attempt to find point sources beyond
this radius.

In regions within the main body of NGC~1569 where the local background due to 
diffuse emission is high, point source detection becomes less robust. To 
restrict our source list to point-like emission regions, we retained only 
those detections with sizes (as defined by the radius encircling 0.8 of the 
total flux) less than 2 times the size of the nominal Chandra PSF (i.e., 
sources with source-size to PSF ratios less than 2.0).  The resulting 45 
point-like sources are listed in Table~1 with their J2000 positions, counts 
extracted from the unsmoothed images, and 1$\sigma$ uncertainties estimated
using Poisson statistics. To avoid neighboring sources, background annuli 
were defined manually around each aperture defined by {\it Celldetect}. 
Within NGC 1569, the point sources detected at the lowest 
significance level all have very hard colors (count rates are highest in the 
H band) which is consistent with the signature of an X-ray binary (or perhaps 
background AGN) seen behind an HI column of few$\times10^{21}~cm^{-2}$ but 
inconsistent with soft, thermal sources. We therefore believe that we have 
successfully culled local maxima in the diffuse emission from the sample
in Table~1.  In \S 3.1 we discuss the identify of the point sources further.


\subsection{Optical Observations}

Narrowband, optical images were obtained at \Ha (6571\angs, 84\angs\ FWHM) and 
in the adjacent continuum (6487\angs, 67\angs\ FWHM) at the KPNO 2.1~m 
telescope
2000 January 2-10.  Fixed pattern noise was removed from the CCD frames in the
standard way using short bias level exposures, dome flats, and twilight
sky flats. The scaled, continuum image was subtracted from the on-band
image to produce an [NII]+\Ha image.  The images were flux calibrated using
observations of spectrophotometric standard stars (Massey \et 1988).  
The \Ha flux was corrected for [NII] 6548,84 emission using the 
luminosity-weighted value of the [NII] 6548,84/\Ha flux ratio, 0.042
(Martin 1998).  We measured the  Balmer decrement along longslit spectra, 
previously described by Martin (1997), and  estimate the luminosity-weighted 
average extinction is $A(\Ha) = 1.48$~mag (for a ratio of general to 
selective extinction $R = 3.1$).  The resulting extinction-corrected \Ha 
luminosity is $5.04 \times 10^{40}$\ergsec.  

In dust-poor galaxies, the \Ha luminosity provides an excellent measure
of the star formation rate (SFR) (e.g., Kennicutt 1998 and references therein).
It is directly proportional to the ionization rate and, hence, the number
of stars with masses greater than about 20\msun.  For a stellar initial
mass function (IMF) with power law index $\alpha = -2.35$ from 1.0\msun\
to 100\msun, the \Ha luminosity implies a SFR of 0.16\msunyr (Leitherer \&
Heckman 1995; Kennicutt 1998). The main uncertainty in
the SFR is the variation in extinction among the HII
regions and warm, diffuse ionized medium  (Kobulnicky \& Skillman 1997;
Martin 1997).  Ionizing radiation absorbed by dust will be re-radiated in
the far infrared, however, so we can check the maximum magnitude of our
error. The IRAS fluxes give a total $L_{FIR}$ of $5.63 \times 10^8$\lsun, 
which is about half of the B-band luminosity of $1.3 \times 10^9$\lsun\ 
(Israel 1988). Their sum $L_{Bol}$ implies a SFR of 0.17\msunyr 
(Meurer \et 1997).  Since the optical and infrared estimates agree, dust
obscuration does not significantly change the estimated SFR.  (Note that
the star formation rate would be 2.5 times higher if the IMF is extended
from 1.0\msun\ to 0.1\msun.)


\subsection{Radio Observations}

We obtained VLA 3.6 cm (8 GHz) radio continuum observations from the
VLA archive database.  Approximately 65 minutes of data were originally
obtained on 21 Aug 1990 in the B configuration using a bandwidth of 50
MHz.  After standard phase and flux calibration in AIPS, we mapped the
field of NGC 1569.  The resulting maps have a beamsize of 1\arcsec\ and
a $1\sigma$ RMS noise level of 0.03 mJy/beam.  We detected 1 source
projected on the disk of \n1569.

As part of a larger radio continuum study of NGC 1569, we obtained 20
cm (1.4 GHz) continuum observations of NGC 1569 for 3 hours on 13 May
2001 in the B configuration using a bandwidth of 50 MHz.  After
standard phase and flux calibration in AIPS, we mapped the field of NGC
1569.  The resulting maps have a beamsize of 5\arcsec\ and a $1\sigma$
RMS noise level of 0.038 mJy/beam.  Only 1 source was detected
toward \n1569.  A recent 20~cm study at higher resolution (Greve \et (2002) 
finds two additional sources besides the VLA source.  Including two 
marginal detections reported in the Greve \et work brings the total
number of radio SNR candidates to 5.

\section{General Properties of the X-Ray Emission from \n1569}

Previous observations with the High Resolution Imager on ROSAT 
show that half of the $\sim 1$~keV flux from \n1569 comes from a diffuse halo
of dimensions $\sim 3\farcm8 \times 2\farcm2$ (Heckman \et 1995, Heckman \et 1995).  
Spectral models fitted to ASCA observations (Tanaka, Inoue, \& Holt 1994; 
Della Ceca \et 1996) firmly establish the presence of a soft, thermal emission 
component which likely emanates from this extended emission 
(Della Ceca \et 1996). In contrast, the nature of the harder, 
$kT = 3 - 4$~keV or $\Gamma \approx 2$, component has not been well 
constrained. The \x luminosity of the halo emission is too bright 
to come entirely from hot, supernovae ejecta; but entrainment of
ambient, interstellar gas in the hot wind can easily boost the luminosity to 
the observed value  (Heckman \et 1995; Martin \& Kennicutt 1996). The location and 
nature (i.e., evaporation or ablation) of this mass loading has not been
established however.  We exploit the spatial resolution of the Chandra
observations to describe (and remove) the point source population,
examine the relative morphology of the cold, warm, and hot gas,
and relate these components to the integrated \x spectrum.


\subsection{Point Sources}

\subsubsection{Sources Associated with \n1569}

Figure~\ref{fig:KNTR-R_XRAY-SM.PLOT} shows a region near the nucleus of
\n1569 with identification numbers from Table~1 near each X-ray source.  
Fourteen point sources lie projected on the main stellar disk of NGC~1569 
(numbers \#14--19 and \#22--29) and likely reside within NGC 1569. 
Figure~\ref{fig:KNTR-F555_XRAY.PLOT} shows the strongest X-ray sources
on the HST WFPC2 F555W image (courtesy of Deidre Hunter; Hunter \et
2000).  None of the X-ray sources coincide with the two most prominent
optical features -- super starclusters ``A'' and ``B'' (Arp \& Sandage
1985).  X-ray sources \#19 and \#22 do coincide with fainter
starclusters designated \#29 and \#34, respectively, by Hunter \et
(2000).  Table~1 includes these probable identifications along with the
F555W magnitudes of the clusters from Hunter \et (2000).  The other 12
X-ray sources which lie projected against the main disk of NGC 1569 do
not have obvious optical counterparts, and approximate upper limits are
listed in Table~1 for their R-band magnitudes.   Four \x sources were
detected in the VLA radio observations, but only source \#28 
appears to be associated with \n1569.  The MERLIN observations
(Greve \et 2002) detect our source \#14 (M-3) as well \#28 (M-6)

Source \#28 has a radio spectral index, $\alpha_3^{20}=-0.9$.
The steep spectrum suggest a non-thermal origin and is typical of evolved
supernova remnants (Weiler \& Sramek 1988).  Optical line imaging in [OIII] 
\lines5007,4959 and [SII] \lines6717,31 reveals a small shell at this location
confirming the identification as a supernova remnant which is particularly
spectacular in the HST imaging presented by Shopbell \et (2000).  The 20 cm 
radio flux of 5.6 mJy implies a spectral luminosity of 
$3.2\times10^{25}~erg~s^{-1}~Hz^{-1}$ which is approximately as luminous as 
Cas A, the most luminous SNR in the Milky
Way  at $3.5\times10^{25}~erg~s^{-1}~Hz^{-1}$
and nearly as luminous as the ultra-luminous SNRs in M~82
($3.2\times10^{25}~erg~s^{-1}~Hz^{-1}$; Allen \& Kronberg 1998).
With 25 photons detected  from 0.3--6 keV (0.0003 ph/s), 
the implied flux is $2.6\times10^{-15}~erg~s^{-1}~cm^{-2}$ for  
a 0.2 keV blackbody and $N(H)=8\times10^{21}~cm^{-2}$. 
The implied unabsorbed luminosity at 2.2 Mpc is $1.4\times10^{36}~erg~s^{-1}$,
comparable to the X-ray luminosities of SNR's in NGC~300 (Pannuti \et 2000) 
or M101 (Pence, Snowden \& Mukai 2001).  


Comparison of the brightest point sources detected with Chandra to
the sources seen in the ROSAT HRI study of Heckman \et\ (1995)
reveals significant variability.  Our Source \#16
corresponds to Source \#1 of Heckman \et\ (1995) and is approximately
3 times more luminous in 2000 May Chandra data than in 1992 March ROSAT data.
Source \#2 of Heckman \et\ (1995) has vanished, and it is not present in
the 2000 Chandra data to the detection limit of $1.2\times10^{-16}$ erg
s$^{-1}$ cm$^{-2}$.  Two of the strongest sources in 2000 May, our \#19 and
\#22, have apparently turned on since 1992 March, since they do not
appear at all in the ROSAT map.  This variability is readily explained if the
sources are accreting X-ray binaries.

To compare the properties of all 14 
point sources projected against \n1569, we measured \x colors
sensitive to absorbing column, $C_1 \equiv (M-S)/(M+S)$, and spectral
hardness, $C_2 \equiv (H-M)/(H+M)$. The top panel of Figure~\ref{fig:2color} 
shows all the sources are harder than the supernova remnant, \#28, so 
most are likely accretion powered binaries.  Their location in
the starburst region indicates they are high mass \x binaries.  Indeed,
since radio-bright SNRs are 
predicted to be older than the X-ray-bright remnants (Chevalier \& Fransson 
2001), a large number of \x SNRs would have been surprising in \n1569.

We extracted spectra of the brightest sources, \#16 and \#19. 
The fitted  emission models, summarized in Table~2
rule out a blackbody emission model but do not distinguish between
a thermal MEKAL model and a power law model. Adopting the power law model 
for consistency with the X-ray binary interpretation, the implied unabsorbed 
luminosity is $3.1\times10^{38}$ erg $s^{-1}$ for source \#16 and 
$5.4\times10^{37}$ erg $s^{-1}$ for source \#19.  
The estimated unabsorbed 0.3--6~keV luminosities for the other sources
range from $1.2\times10^{35}~erg~s^{-1}$ to $1\times10^{37}~erg~s^{-1}$,
where we have assumed a mean photon index of $\Gamma=3.0$ and a total 
foreground column of $0.21\times10^{22}~cm^{-2}$ (see   
Figure~\ref{fig:2color}).

The models in Figure~\ref{fig:2color} demonstrate that
the point sources require foreground absorption columns 
of 0.6 to 1.6 $\times10^{22}~cm^{-2}$, {\it in addition to} the 
Galactic foreground of N(H)=$0.21\times10^{22}~cm^{-2}$.  
Using a 21-cm map
kindly provided by E. Wilcots (pvt. comm), we found the inferred \x 
absorption columns 
toward individual point sources are loosely correlated with the 21-cm intensity
in the same direction.  Most notable are the two brightest sources which
lie in a previously described HI hole (Israel \& van Driel 1990).







Figure~\ref{fig:hardl_hist} shows the cumulative luminosity function
of the \n1569 point sources in the hard band (1.1--6 keV).  With only 14 
data points, we do not attempt to fit the slope. The solid line simply
illustrates a typical power law index $\alpha=0.5$, where $S(>C)=KC^{-\alpha}$,
for the point source population in star-forming galaxies (Prestwich 2002; 
Tennant \et 2001).  We have normalized it to the highest luminosity bin which 
contains 5 sources with a total of 1400 counts in the 1.1--6.0 keV band.  This
luminosity function predicts a total of 1800 counts in the hard band,
only 18\% more than the observed total of $\sim$1520 in the 14
identified point sources.    If the slope of the luminosity function
were steeper, say $\alpha=0.9$, then the total predicted contribution from
unresolved sources in the hard band would be 6300 counts, more
than 4 times the total observed.  However, this slope would predict 400
hard sources with more than 10 counts in the 85 ksec integration. Given that 
we have detected point sources to a completeness level of 
essentially 100\% for sources $>$10 counts, the fact that we only detect 9 
sources with more than 10 counts in the hard band constrains the hard X-ray 
luminosity function to values $\alpha\leq0.55$. 
In conclusion, 1) the slope, $\alpha$, of the cumulative hard X-ray
luminosity function,  is constrained to be
$\alpha\sim$0.5 and 2) the total hard counts from unresolved X-ray
point sources do not contribute more than 18\% above the contribution
from the 14 currently-identified point sources.
The diffuse emission can therefore be studied with little contamination from
the  point source population by filtering out the cataloged point sources.

\subsubsection{Other Sources Projected Near \n1569}


Of the 31 X-ray sources which lie outside of the main body of NGC~1569,
about half coincide with point-like optical counterparts and are
probably foreground Galactic stars.  Others may be background AGN.  We
make tentative identifications by considering the ratios of
X-ray to optical flux, $f_X/f_V$.  Table~1 lists this ratio for each
source.  Based on Figures 2 and 3 of Krautter \et (1999), we tentatively 
identify objects with $f_X/f_V>-0.5$ as stars, and objects with $f_X/f_V<-0.5$
as probable AGN.  Using this criterion, 15 objects are probable AGN and 16 are
stars.  Given the $2\sigma$ sensitivity limit of
$\sim4\times10^{-16}~erg^{-1}~cm^{-2}$ in the broadband 0.3--6
keV  image, we would expect 16$\pm$4 extragalactic background sources 
over the 7\arcmin\ x 7\arcmin\ area analyzed on the S3 chip, based on
the Chandra Deep Field South count rates (Rosati \et 2001).
This is consistent with the 15 objects we have tentatively identified as
background AGN.

The lower panel of Figure~\ref{fig:2color} shows the 31 sources
which lie outside \n1569, coded by their identification as either Star or
AGN.  Lines denote MEKAL thermal plasmas with temperatures kT=1.0 keV and
kT=2.0 keV for five different foreground HI column densities as
labeled.  Objects identified as probable stars in Table~1 on the basis
of $f_X/f_V$ occupy a broad locus with lower implied HI column density
than the probable AGN which cluster more tightly and have higher
implied foreground column density.  This separation provides an
independant way to separate probable foreground stars from extragalactic
objects.




\subsection{Spatially Extended X-Ray Emission}

\subsubsection{X-Ray Morphology}

\fig~\ref{fig:all2_ha} shows the \x surface brightness contours
around \n1569.
The halo emission consists of two pairs of lobes at position 
angle 50\deg (NE and SW lobes) and 160\deg (NW and SE lobes) which were 
previously detected in the ROSAT imaging (Heckman \et 1995; Della Ceca \et 1996).
The signal-to-noise
 ratio in the northeast lobe, at a projected distance of 1.682~kpc, is 
$5\sigma$ and $7\sigma$ in the soft and medium bands, respectively.  The 
detection level of the southwest lobe, at 1.912~kpc, is $2\sigma$. 
Two point sources (\#30 and \#31), not believed associated with the galaxy, 
contribute to the emission from the southeast lobe.  In contrast to the
halo contours, the higher surface brightness
contours in \fig~\ref{fig:all2_ha} lie within 40\asec 
of the major axis and follow the ellipticity of the optical disk. This
change in the morphology of the contours suggests that at least two
physical components of the galaxy contribute to the diffuse emission.
It is also interesting to note that the intensity gradient is steepest toward 
the west edge of the disk where molecular clouds have been detected (e.g., 
Taylor \et 1999).  The shock that is heating the hot disk gas is likely
encountering the densest regions of the disk there.

Heckman \et (1995) and Della Ceca \et (1996) previously noted that the 
position angle of the lobes matches that of the most extended \Ha 
emission. The nature of this correspondence can now be seen more clearly in 
\fig~\ref{fig:all2_ha} and \fig~\ref{fig:all2_ha_inset} as well as 
in the color image of Figure~\ref{fig:haxr}. The 
northeast lobe is bounded by filaments 12 and 13 (names from Hunter \et 1993).
Filament 11 marks the western edge of the northwest lobe.  The \x surface 
brightness falls off abruptly where the southwest lobe meets the bright 
\Ha {\it arm}, and Filament 10 marks the extent of the hot gas further above 
the disk plane. To the southeast, the intensity of the halo emission declines 
sharply beyond Filament 7,  but Filament 8 follows the most extended emission.
In Figure~\ref{fig:haxr}, much of the periphery of the \x nebula (green) is 
marked by faint \Ha filaments (red). Hence, the \x emission is detected over 
the full extent of the \Ha nebula but not convincingly beyond it.

The location of the cooler (i.e., \Ha emitting) gas is therefore consistent 
with a web of filaments surrounding the hot halo.  In the \n253 superwind, 
the \x emission appears to come mainly from a boundary layer adjacent to
the shell rather the entire interior of the bubble (Strickland \et 2000).
In contrast to the cone-like geometry of the \n253 wind, 
the hot halo gas in \n1569 emanates from nearly the entire 
disk and maintains roughly the same diameter to distances at least 2~kpc 
above the midplane of the disk, whose position is indicated  by the 21~cm 
contours in \fig~\ref{fig:haxr}).   Much of the disk plane is clearly 
filled with hot gas, and the \x lobes appear to fill a large fraction of a 
cylindrical volume of radius similar to that of the star-forming disk.  The 
\x halo is believed to be limb brightened in the southwest arm but appears 
more diffuse in the other lobes.

The cavities of the bubbles are easier to identify kinematically, since
the limbs of the shells show no Doppler shift.  The location of local maxima 
in the line-of-sight expansion speed (measured from the \Ha emission)
are marked by the letters (from Martin 1998) in \fig~\ref{fig:all2_ha_inset}.
Near the base of Shell~A and 
Shell~E, the \x contours poke outwards into the lobes at PA=50\deg.  The NW 
lobe also appears to fill the region defined kinematically by Shell~B. 
Somewhat brighter contours, about 30\asec from the midplane,  protrude into 
the expanding cavities that define Shell~D and Shell~G.  This correlation 
between the \x intensity and the kinematics of the warm gas suggests
some of the \x emission comes from near the center of the bubbles.
The exception is the strong intensity enhancement just inside the 
southwest arm which we attributed to boundary layer emission.


\subsubsection{X-Ray Color}

Figure~\ref{fig:3x} shows the composite color \x image obtained from
our Soft (red), Medium (green), and Hard (blue) images.  The spatio-spectral
information immediately isolates the dominate spectral components.  First,
the hardest emission (blue) comes from the  point sources which are 
concentrated toward the center of the disk.  Second, a strong color gradient 
is visible across the galactic disk, perpendicular to the major axis.
The emission is harder (more blue) north of the midplane.
Third, in the southern halo, the $\sim 1$~keV emission (green) is prominent
along the inner (i.e., concave) side of Filament~6, near Filaments 11 and 10, 
and around Filament~5. In contrast, the softest halo emission (red) 
is not correlated with bright \Ha features. 
The softest regions in the NE lobe are also well inside the \Ha filaments.
The point sources were shown to be hard sources in a previous section, and
in \S~\ref{sec:disk_abs} we show quantitatively that photoelectric absorption 
from the inclined gas disk produces the color gradient across the disk.

The surprising characteristic of Figure~\ref{fig:3x} is the color variation 
in the halo.  The halo does not present a simple radial spectral softening as 
would be expected from a free flowing wind cooling adiabatically (e.g.,
Chevalier \& Clegg 1985). The emission is strongest in the Medium band (green 
regions) near the bright \Ha filaments.  The Soft band (red regions), in 
contrast, become more prominent between the filaments. The location of 
the softest regions in the \x color $C_1$ are denoted by the minima of the
contours in \fig~\ref{fig:mscon_x}.  The regions with the softest color,
most visible in the southern halo,  clearly emit the lowest total 
\x intensity. In contrast, the high surface brightness regions have
a {\it green} color in \fig~\ref{fig:3x}, emit most of the halo luminosity,
and must be denser regions of the plasma. Any physical model must explain 
why the brighter, denser regions are located near the \Ha filaments.  

To produce the observed \x luminosities, a
wind must be confined by a dense shell of halo gas or loaded with
entrained interstellar gas (e.g., Suchkov \et 1996).  A shocked shell
could produce the observed limb brightening.  Alternatively, the mixing 
layer of intermediate temperature gas between the shock and the wind fluid
could generate most of \x emission. Unlike \n253, where all the halo
emission in the Chandra observation is attributed to such mixing layers
(Strickland \et 2000), \fig~\ref{fig:3x} also reveals the lower surface 
brightness wind fluid -- i.e., the {\it redder} regions.

Our spectral fitting, see \S 3.3.3, constrains only 
the temperature of the bright, limb-brightened component. The soft emission 
is too faint for spectral analysis. We can compare the colors of these
regions, however. The colors of the red regions are $C_1 = (M-S)/ (M+S) = 
0.36 \pm 0.07$ and $C_2 = (H-M)/ (H+M) = -0.29 \pm 0.07$. The green region is 
harder in the soft color, $C_1 = 0.65 \pm 0.05$, but softer in the hard color, 
$C_2 = -0.41 \pm 0.06$.  The bright halo regions (green) are
harder than the faint (red) regions in the color $C_1 = (M-S)/(M+S)$, so
they are likely more absorbed.  Comparison to the colors of the equilibrium, 
thermal emission models in \fig~\ref{fig:2color}b indicates the absorbing 
column could be as much as $1 \times 10^{21}$\col\ higher toward the limb 
brightened regions. This amount of differential attenuation seems plausible
based on the  spatial variations shown in deep, 21~cm map of the halo
(M\"uhle, 2002, pvt. commn.). Comparison of our $C_1$ color map in
\fig~\ref{fig:mscon_x} to her HI map on a point by point basis did
not yield a simple correlation, but further investigation may clarify
the relation of  the $C_1$ to the absorption pattern.

The bright halo regions (green) are actually marginally
softer than the {\it red} regions in the color $C_2$. The
red regions may include hotter gas. Alternatively, they could have a higher 
abundance of alpha elements relative to Fe. The Fe~L emission lines contribute 
much of the flux in our Medium band, while strong lines from alpha-process
elements are present in both the Soft band (OVII 0.574~keV, OVIII 0.653~keV)
and Hard band (Mg XI 1.34, 1.35~keV, Si XIII 1.87~keV).  



\subsection{The Integrated Spectrum}
			
Figure~\ref{fig:int} shows the integrated \x spectrum.\footnote{
 	The integrated spectrum was extracted from an elliptical
	aperture enclosing the galaxy at 67.706143\deg, 64.844363\deg 
	(J2000) with 
	position angle 30\deg, major axis 322\asec, and axis ratio of 0.582.}
Absorption from the Galactic halo causes much of the soft energy turnover.
We illustrate the contribution of the point sources previously identified
\S~\ref{sec:detect}.  Their contribution is clearly harder than the total 
spectrum and shows no emission lines. The remaining diffuse emission presents 
strong lines from OVII 0.57~keV, OVIII 0.65~keV, NeIX 0.905, 0.922, 0.915~keV,
NeX 1.1016~keV, MgXI 1.34, 1.35~keV, and SiXIII 1.87~keV. 
The count rate peaks just below $1~keV$ where Fe L-shell lines, most
prominently FeXVII,  are blended at the energy resolution of the Chandra CCDs
(a few hundred eV). Note that the line spectrum is not sensitive to the
details of the background subtraction.\footnote{
	We use the background extracted from the data at large radii, and 
	corrected for effective area.	The background in the data has a 
	slightly higher normalization at low energy, $E < 0.55$~keV, than 
	that obtained from the simulated background.
	The difference, a factor of 1.7, is consistent with the spectral 
	variabilty among pointings in the data sets used for the background 
	simulation (Markevitch, pvt. comm.).}
The spurious feature at 1.6~keV in both the data and the background is thought
to be an artifact of the gain tables.


The energy distribution of the counts from the diffuse emission is quite
broad compared to the folded spectrum of any single-temperature thermal 
emission model.  (See the fit statistic of the single-temperature model in 
Table~3!)  By suppressing the emission lines, the fitting routine can 
increase the normalization which broadens the count distribution.  When
we allowed the metallicity to be driven to the unphysically low 
value of 0.06\zsun, the fit statistic was improved to $\chi_{\nu^2} =
1.55$, which is still unacceptable.  We therefore conclude that
the diffuse emission is inherently multi-temperature.

\subsubsection{Comparison to Previous Spectral Models}

Della-Ceca \et (1996) excluded single-temperature thermal models on the same 
basis in their analysis of the joint ASCA and ROSAT/PSPC datasets for \n1569.
They suggested two components, representing the disk and halo emission, are 
required and discussed four such models (RS+RS, M+M, RS+Po, and M+Po). The red 
curve in Figure~\ref{fig:ellip_fit} shows the residuals from their M+M model 
folded through the Chandra response.  The Chandra data require a 17\% higher 
normalization.  The fit statistics in Table~\ref{tab:int}
do not distinguish the M+M and M+Po models. The MEKAL models fit slightly 
better than the Raymond-Smith models at low energy, and we use them for 
discussion in the rest of our analysis.  The fitted temperatures are slightly 
higher for the Raymond-Smith models.

The residuals of the renormalized Della Ceca \et (1996) M+M model are shown in the bottom panel of Figure~\ref{fig:ellip_fit}. 
Della Ceca \et (1996) attributed the residuals around 0.6-0.7~keV (near the oxygen lines) to 
calibration problems.  Although the calibration of the Chandra response
below $\sim 0.7$~keV is not firmly established, the reappearance of the
residuals in data from a different observatory suggests they are real.
More importantly though, the Della Ceca \et (1996) M+M model fails to reproduce the strong 
Mg XI lines seen in the spectrum at $\sim 1.3$~keV.  The residuals near the 
NeIX and NeX complex are also significant.  Taken together, these shortcomings
indicate the line emisson of the model is too weak.  We can use the spatial
information afforded by the Chandra data to construct a better model
of the spectrum.

For consistency with previous analyses of ROSAT and ASCA observations of \n1569 
(Heckman \et 1995; Della Ceca \et 1996), we adopt $\nhg = 2.1 \times 
10^{21}$\col; but the foreground column could be up to 20\% lower. 
The galaxy is projected only 11\deg above the Galactic plane.  Measurement
of the Galactic \nh column is complicated by the proximity of emission 
from \n1569 in velocity space ($v_{sys} =  -77$\kms and 119\kms FWZP,
Reakes 1980). Averaged over a scale of 30\amin, the total HI column 
toward \n1569 is $2.1 \times 10^{21}$\col (Burton 1985). However, the average
of eight profiles extracted 30\amin away from \n1569 gives a foreground 
column of $\nhg = (1.93 \pm 0.16) \times 10^{21}$\col.\footnote{  
	The 21-cm $T_A(v)$ profile toward \n1569 shows five local maxima 
	between -150\kms and 30\kms. If only the emission at $v < 30$\kms was 
	attributed to the foreground component, the absorbing column is reduced 
	to $\nhg = 1.7 \times 10^{21}$\col.]}

\subsubsection{Revised Spectral Components}

The strong \x color variations described above led us to suspect that the 
absorbing columns of the two spectral components were different.  
\fig~\ref{fig:ellip_fit} shows another model (blue line) in which 
the intrinsic absorption of each thermal component was allowed to vary 
independently.  It fits the count 
distribution around MgXI better as shown by the residuals in the bottom panel 
and the fit statistics in Table~\ref{tab:int}.  The harder, 0.7~keV, component 
is several times more absorbed than the soft, 0.3~keV, component.   In this
model, the Mg line becomes stronger (and better describes the data)
because the extra absorption allows for a lower temperature for the hard
component than Della Ceca \et derived.  Both thermal components contribute
significantly to the Mg line emission in our revised model.

We introduce a third emission component to describe the contribution of
the \x point sources to the integrated spectrum.  We fitted the summed
spectra of the point sources, blue line in \fig~\ref{fig:int}, with
a single power law model since it  shows no prominent line emission. 
The point source population is adequately described by an absorbing column
of $N_H = 2.3 \times 10^{21}\col$, power law index $\Gamma = 2.40$, and 
normalization of $8.40 \times
10^{-5}$~photons keV$^{-1}$~cm$^{-2}$~s$^{-1}$~(at 1~kev).  (This absorption 
is in addition to the foreground column of $2.1 \times 10^{21}$\col\ and
has an assumed metallicity of 0.25\zsun). Since the normalization of the 
power law component is quite low,  a description of the integrated spectrum 
still requires two components to produce the wide spread in energy exhibited
by the counts as shown by the final model in Table~\ref{tab:int}.


In summary, in contrast to Della Ceca \et (1996), we find strong evidence for absorption 
intrinsic to \n1569.  Including this intrinsic absorption in the spectral
model drives the temperatures fitted to the two thermal components lower.
The fraction of the flux coming from point sources increases toward
higher energies.  In our Hard band, point sources contributes 57\% of 
1.1 to 6 keV flux. The
thermal components emit strong lines (e.g., MgXI 1.343,1.352~keV and 
SiXIII 1.87~keV) in the Hard band. The 0.25~keV thermal component emits 72\% 
of the Soft band flux.  The total 
0.3--6~keV flux is $4.11 \times 10^{-13}$\flux, and the unabsorbed 
luminosity in this band is $1.02 \times 10^{30}$\ergsec. 

The shortcomings of this {\it wabs*pow + wabs*mek + wabs*mek} description
are the remaining line residuals, most noticeably around MgXI.  One way to 
boost the MgXI line is to add a third thermal component with a high 
normalization and a high intrinsic absorption component. Columns of order 
$N_H \sim 10^{22}\col$ would be required, however, to avoid overproducing the 
OVII and OVIII line emission.   This solution seems unlikely because the 
galaxy contains no known gas component that would cause such extreme 
absorption in the halo. Such high columns, if present, would have to 
be confined to small regions unresolved by the 21~cm observations; but
the MgXI line emission seems to come from a large area. To find acceptable 
fits to the lines, we vary the element abundances as described in \S 4.2.


\subsubsection{Fits to Southern Halo Spetrum}

To better understand the composite nature of the southern outflow,
we extracted spectra  from the {\it red} and {\it green} regions
in \fig~\ref{fig:3x} separately.  (Recall that the green regions have higher
\x surface brightness and are found adjacent to the brightest \Ha filaments
to the south of the disk, while the red regions are  spatially correlated with 
the centers of the bubbles.) The count distributions in \fig~\ref{fig:rg_spec} 
illustrate the deficit of 0.7 to 1.1~keV counts in the {\it red} regions 
relative to the {\it green} regions.  A MEKAL model with photoelectric 
absorption provides an excellent description of the {\it green} regions.  The 
best fit parameters are $kT = 0.30$~keV, $N_H = 3.4 \times 10^{21}\col$,  and
$norm = 8.9 \times 10^{-5}$\norm.  The reduced chi-squared value is 0.93, and
the iron group abundance is 0.48 when the alpha elements are fixed at
solar abundance. The bright regions in the southern halo with the 
{\it green} color are clearly the origin of the soft, 0.3~keV, component
in the integrated spectrum.  

The poor quality of the {\it red} spectrum (about 400 counts) does not 
allow us to distinguish between a model of a hotter plasma with minimal
absorption and one with a very high alpha element to iron abundance ratio.
However, for purposes of illustration, we extracted a spectrum of the diffuse 
emission south of the galactic midplane (apertures S1-5) and fitted a two 
temperature model.  Table~\ref{tab:s1-5} summarizes the results.  Like our
model for the integrated spectrum, models of the emission from the
southern half of the galaxy also require two thermal components.
The only change to the 0.7 keV component is a decrease in the normalization
which is consistent with removing half of the volume from the aperture.
The fitted absorbing column for 0.3~keV component is also significantly lower,
which probably reflects the relative geometry of the two lobes.
These results support  the physical association of the hotter
thermal component with the flow in the disk and the cooler one with the halo.
The second halo component is too weak to  affect the integrated spectrum much.

%

\section{X-Ray Measurements of Metallicity}

Each component added to the model of the integrated spectrum was
spatially identified in the spectral-imaging data.   Such spectral 
decomposition is a necessary prerequisite for fitting abundances.
Otherwise, several degeneracies (e.g., between metallicity and temperature, 
emission measure, and absorbing column) can lead to unphysical, yet 
statistically acceptable, solutions for the abundances.

Before using our spectral model to investigate the metallicity of
the hot wind, we examine one of the new spectral components --
the disk absorption -- in more detail.  Since the photoelectric
absorption cross section is dominated by metals at energies greater than 
$0.25~keV$ (Morrison \& McCammon 1983), the equivalent hydrogen column is 
sensitive to the assumed metal abundance of the disk gas. We fit the \x 
absorption column and compare it to optical and radio measurements 
of the H column to constrain the metallicity of the gas disk. 

\subsection{Disk Absorption and Metallicity} \label{sec:disk_abs}

Figure~\ref{fig:geometry} shows the relevant geometry of the disk and wind. 
North (south) of the major axis, the gas kinematics indicate the wind is 
pointed away from (toward) our sightline (Israel \& van Driel 1990; 
Heckman \et 1995; Martin 1999). The prominent change in \x color perpendicular 
to the major axis of the disk, i.e., the emission is harder on the north side 
of the midplane, is consistent with the northern lobe of the outflow being 
viewed through the disk because lower energy X-rays are most easily absorbed. 

To measure the projected gradient in the absorbing column, we assumed that
SSC~A is near the dynamical center of the disk (Israel \& van Driel 1990;
Stil \& Israel 1999) and extracted spectra at 20\asec\ (213 pc) intervals 
along the minor axis as illustrated in \fig~\ref{fig:all2_ha_inset}.
Figure~\ref{fig:ns} shows the deficit of soft counts in the northern spectrum 
relative to the southern spectrum for several pairs of apertures. For
comparison, the spectra are normalized at high energies since the summed 
northern spectrum (N1-5) has $\sim 50\%$ fewer counts than the summed 
southern spectrum (S1-5).  For a solar metallicity disk, the N1-5 spectrum 
requires an an additional absorbing column of, $\Delta N_H \approx 1.6 \times 
10^{21}\col$ relative to the column toward S1-5 spectrum.  
Comparison to the S1 and N1 spectra in \fig~\ref{fig:ns} indicates that
this difference comes largely from emission at projected heights of 0
to 20\asec (aperture 1). 

We also fitted spectra from apertures S1, N1 and N2, which are marked in 
\fig~\ref{fig:all2_ha_inset}, individually. A thermal emission 
model (MEKAL, 0.25\zsun) with variable photoelectric absorption and a fixed 
Galactic  column (solar metallicity) was used in each case.  
Figure~\ref{fig:zH} shows the 
variation in absorbing column with projected height above the disk for three 
different values of the disk metallicity.  To the north, the fitted absorbing 
column drops with projected distance from the midplane.  The gradient is 
measured out to a projected midplane distance of 40\asec in absorption, which 
corresponds to galactocentric radii of $\sim 426$~pc.  South of the disk 
midplane, the fitted column at point S1 is lower than that at N1 but
does not drop all the way to the foreground value. The scale height of the 
absorbing gas must be $\sgreat\ 123$~pc to produce this absorbing column
south of the midplane.

The projected N(HI) column measured in 21-cm emission (Israel \& van Driel 
1998) shows a similar gradient as illustrated by the solid black line in 
\fig~\ref{fig:zH}.
The equivalent hydrogen columns inferred for the solar metallicity disk model 
are all significantly lower than the HI emission column.  To match the 
measured hydrogen column, the disk metallicity must be less than solar. 
The green points in \fig~\ref{fig:zH} show that reducing the disk metallicity 
to 0.25\zsun\ produces fitted \x absorption columns consistent with the 21-cm 
emission data.  If the disk metallicity is further reduced to 0.1\zsun, 
the inferred \x column is inconsistent with the 21-cm HI column. 

Absorption from metals in the warm, ionized disk could drive the fitted 
metallicity lower (shifts the black curve further to the right in 
\fig~\ref{fig:zH}).  We therefore estimated the  intrinsic and foreground
ionized gas columns from their \Ha emission measure.  For a volume filling 
factor of 1\%, the average column toward the \Ha nebula (R = 2 kpc) is $2.8 
\times 10^{19} (\epsilon/0.01)^{0.5}$\col. The foreground Galactic column 
estimated from WHAM data assuming a pathlength of 1~kpc is $(3-6) \times 
10^{19} (\epsilon/0.01)^{0.5}$\col\ (Reynolds \et 1998).   Both of these 
values are small compared to the HI columns.  We therefore conclude that the 
mean metallicity of the disk gas is less than solar, greater than 0.1\zsun, 
and consistent with that of the HII regions (0.20 to 0.25\zsun).



\subsection{Metallicity of the Hot Outflow} \label{sec:zwind}

To identify the major spectral components, we assumed a metallicity of 
0.25\zsun, consistent with the HII region oxygen abundance, for the thermal 
plasma.  Using the integrated spectum, we investigate 
the best fitting abundance and the abundance ratios. Our model for the 
integrated count 
distribution includes the following components: (1) a foreground Galactic 
column with solar abundances and $N_H = 2.1 \times 10^{21}\col$; (2) a power 
law component with an intrinsic absorption component of $1.5 \times 
10^{21}\col$, photon index $\Gamma = 2.47$, and normalization $9.20 \times 
10^{-5}$~ph keV$^{-1}$\col s$^{-1}$ at 1~keV; (3) an absorbed thermal 
component described by a {\it MEKAL} model with variable abundances; and (4) 
a hotter, absorbed thermal component with variable abundances.  The abundances
of Mg, Ne, Si, and Ca, relative to their solar values, are tied to the
O abundance ratio; and we refer to this set of elements as the {\it alpha 
elements}, or $Z_{\alpha}$.  Aside from He, which is left fixed at  solar 
abundance, all the other elements are varied with the Fe abundance,
hereafter $Z_{Fe}$.\footnote{
	The solar abundance ratios in XSPEC 10.0 are photospheric values from
	Anders \& Grevesse (1989).  This meteoritic scale from 
	Anders \& Grevesse (1989) has significantly lower iron abundance.
	This controversy does not affect our analysis since we work in ratios 
	relative to solar and define solar consistently throughout. 
	We define solar abundances by the meteoritic values tabulated
	by Anders \& Grevesse (1989).
	The solar values are still undergoing significant
	revision (e.g., Grevesse \et 1996; Prieto \et 2001).  
	}
 
For alpha element abundances from $Z_{\alpha} =  0.1$\zsun\ to 100\zsun,  
we fitted the temperature, absorbing column, and normalization of 
the two thermal components. The fit statistics in Table~\ref{tab:alpfe} show
that the best fit is obtained for $Z_{\alpha} = 1.0 \zsun$. The fit is poor if 
$Z_{\alpha} \le\ 0.25\zsun$, but abundances higher than solar cannot be 
excluded.  Higher metallicities are simply compensated by lower emission 
measures in the spectral fitting.  Figure~\ref{fig:ac} illustrates the 
degeneracy using a model of the integrated spectrum. At temperatures of 0.3 
and 0.7~keV, the spectrum is completely dominated by emission lines which are 
largely blended together at $\sim 100$~eV resolution.  The strength of the 
Fe lines relative to the NeIX, NeX, MgXI, and SiXIII lines largely determines 
the shape of the count distribution.   The spectrum looks almost the same 
when the continuum (i.e., normalization) is suppressed and the line emission 
(i.e., metallicity) is increased.  The exception is at very low metallicity.
The continuum of the $Z_{\alpha} = 0.25\zsun$ model in Figure~\ref{fig:alpfe}, for 
example, cannot be fit between 0.5~keV and 1.2~keV. Hence, we rule
out metallicities lower than 0.25\zsun\ for the thermal emission component. 
A practical upper limit can be placed on the wind metallicity
when the minimum wind mass is considered (see \S 5.2.1).


The ratio $Z_{\alpha} / Z_{Fe}$, in contrast, is well constrained.  
\fig~\ref{fig:amod2_adat2} illustrates the distinct contribution of the 
alpha element emission lines and the iron emission lines to the spectrum. 
We can set the alpha element abundances at just about any level, and a fit is 
found with an iron abundance (relative to solar) less than half the 
alpha element abundance.  For example, the best fitting model,  
$Z_{\alpha} = 1.0 \zsun$, required
$Z_{Fe} = 0.37\zsun$; but  $Z_{Fe}$ increases to 43\zsun\ when $Z_{\alpha}$ 
is set to 100\zsun.  Figure~\ref{fig:con93} shows the 68\%, 90\%, and 99\% 
confidence contours in the $Z_{Fe}$ versus $Z_{\alpha}$ plane.  The
measured value of $Z_{\alpha} / Z_{Fe}$ clearly lies between 2.5 and 4.8
$Z_{\alpha,\odot}/Z_{Fe,\odot}$.  Normalizing to the meteortic AG89 solar
abundances decreases this ratio by 0.086~dex to $Z_{\alpha} / Z_{Fe}$  
between 2.1 and 3.9.

In principle, the alpha element to iron abundance ratio constrains the origin 
of the metals. To illustrate the argument, we adopt the 0.1\zsun\ series A 
models of Woosely \& Weaver (1995). The IMF-averaged supernova ejecta have an 
O abundance 8~$Z_{O,\odot}$ and an Fe abundance of 0.2~$Z_{Fe,\odot}$.
If the ejecta are mixed with interstellar gas of average metallicity 0.2\zsun,
then the range of mass-loading 
factors, $\chi \equiv M_x / M_{ej} -1$, implied by the measured alpha element 
enrichment is $\chi = 12$ to 36.  These values should not be taken as exact
limits because yield calculations for Type~II supernovae differ by up to a 
factor of three  (e.g., Gibson, Loewenstein, \& Mushotzky 1997 and references
therein).  The measured ratio of $Z_{\alpha}/Z_{Fe}$ in the outflow
is supersolar but much lower than the ratio predicted for pure 
supernova ejecta.  We note that the alpha element to iron abundance ratios
reported for halo stars are similar over a broad sub-solar metallicity range 
(c.f., Wheeler \et 1989 Figures 2 and 3).





%

\subsection{Limits on Other Hot Reservoirs of Metals} \label{sec:other}

%

\subsubsection{Temperature Range in the Wind}

Attempts to measure temperature variations with scale height from the S1 
through S5 spectra yielded no clear trend.  The color \x image in
\fig~\ref{fig:3x} shows why.  The color variations are not  
radial as would be expected in a freely expanding wind which cools 
adiabatically.   Furthermore, as noted above, it is not clear that the
red and green regions in the halo actually have significantly different
temperatures.  The red regions are less absorbed and appear to have
a broader temperature range, but the number of photons is not high
enough to describe the temperature range accurately.

The lines identified in the spectrum are consistent with the ionization
balance expected in collisional ionization equilibrium for the
fitted temperatures of the thermal components, 0.3 and 0.7~keV.
Both components contribute to most of the lines.  As can be seen in
\fig~\ref{fig:ac}, the exception is OVII, which is produced almost
entirely by the 0.3~keV component.  MgXII and SiXIV are present as
a minority species in the hotter component alone, but their lines are
too weak to detect.  Indeed the absence of these lines in the integrated
spectrum, see \fig~\ref{fig:int}, is the strongest constraint on the emission 
measure of gas
significantly hotter than 0.7~keV. Higher temperature
gas would require an extremely low emission measure to avoid detection
via these lines.  Cooler gas, on the other hand, is plausibly (likely)
present; but the high Galactic HI column in the direction of \n1569
combined with the uncertainties in the Chandra calibration at low
energy prevent us from establishing any useful constraint on the cooler
gas.

Strickland \et (2000) have emphasized that the hottest component of
the outflow may be very difficult to detect. We place upper
limits on the normalization of components at 1.82~keV
and 4.56~keV of $1.36 \times 10^{-5}$\norm and $1.51 \times 10^{-5}$\norm
respectively. (These are 68\% confidence limits.)  Hence the
surface brightness of a hotter component is down by a factor of
at least 100 relative to the 0.3~keV and 0.7~keV components.



\section{Discussion}

The chemical and dynamical evolution of the wind are coupled
because the supernova explosions that power the galactic outflow
also enrich it with metals.  Hydrodynamical models suggest that most of 
the material carried by a galactic wind is entrained interstellar
gas rather than pure supernova ejecta (e.g., Suchkov et al 1996). 
This mass loading 
provides a second coupling between the dynamical and chemical evolution of 
the galactic wind since it reduces the temperature of the hot, 
supernova-driven wind while diluting the wind metallicity.   By requiring
the dynamical and chemical models to be self-consistent, we can constrain
the properties of the outflow much better than was possible previously.
The results allow us to discuss the role of the wind in the chemical
evolution of the galaxy.

\subsection{Wind Dynamics}

\subsubsection{The Existing Picture}

The bipolar outflow in \n1569 was first recognized in \Ha images (Hodge 1974;
deVaucoulers et al. 1974).  Studies of the stellar population and radio continuum
concluded the starburst was 10-20~Myr old (Israel ; O'Connell; Hunter \et 1993)
consistent with the measured dynamical age of the outflow (Heckman \et 1995; Martin 1998).
The mechanical energy produced, see Table~\ref{tab:sb}, is equivalent to
3000 to 26,000 supernova explosions depending on whether the full starburst population 
or just the two super star clusters are considered.  In either scenario, the implied 
rate of energy injection exceeds that required to generate a superbubble which 
breaks out of the disk supersonically (e.g., Martin 1998).

The exact power requirement of the outflow in \n1569 is sensitive to the assumed 
distribution of disk and halo gas. For an average ambient density $n_0$,
the total mechanical energy needed to reproduce the observed sizes and expansion 
speeds of the shells is  $\sim 2.8 \times 10^{55} (n_0/0.1~{\rm cm}^{-3})$~erg, 
where we have added the shells identified in Martin (1998). Comparison to 
Table~\ref{tab:sb} shows that the SSC's alone do not supply enough energy to 
drive the outflow.  The Chandra images support the conclusion that the full
starburst population powers the outflow because the wind appears to emanate
from the entire stellar disk not just the central 100~pc.

Similarly, the Chandra results are consistent with the previous conclusion
that the hot wind is not gravitationally bound to the galaxy (Heckman \et 1995; Martin 1999).
(Note, that in contrast, the warm shell is marginally bound to the dark halo.)
The halo temperature fitted to ROSAT and ASCA data was shown to exceed
the specific thermal energy needed to escape from the halo  (Martin 1999, Figure~3), 
and the same is true of the revised thermal components at 0.3~keV and 0.7~keV. 
The hot wind will therefore escape unless mass loading accelerates the cooling
by a large factor and/or the wind does a great deal of work against a putative
gaseous halo. Since the combined luminosity of the two thermal components in our model,
$L_x(0.3-6~keV) = 8.18 \times 10^{38}$\ergsec, is only a few percent of the average rate 
of mechanical energy injection, the radiative losses from the hot bubble are likely
small. We discuss below whether the surprisingly complex spectrum 
and spatial distribution of the \x halo emission provide indirect evidence for the
existence of such a halo via the interaction of the outflow with it. 

The mass of the hot wind depends on the fitted emission measure and the assumed 
volume filling factor.  Our results also confirm that the \x emission comes from 
the region interior to the \Ha-emitting shells as previous ROSAT imaging indicated
(Heckman \et 1995; Della Ceca \et 1996).  This result supports the assumption that the
volume filling factor of the hot wind is of order unity, $f \sim 1$. If 
the hot wind and warm shell have similar pressure, then the volume filling 
factor of the warm gas must be $\epsilon \approx 0.01$ (Martin 1999).
For these parameters, the implied mass loss rate in the hot wind is a few
tenths of a solar mass per year, which is similar to the star formation
rate  (Martin 1999).  We re-examine the mass carried by the wind, however, 
in light of its dependence on the wind metallicity which we have constrained
for the first time.  We find the rough scaling between mass flux in the wind
and star formation rate still holds for a solar metallicity wind.


\subsubsection{Revised Mass of \X Emitting Nebula and Mass Loading}

In \S~\ref{sec:zwind}, we associated the 0.7~keV and 0.3~keV spectral
components with the thermal emission from the outflow in the disk and
halo, respectively. The fitted emission measures of these thermal components 
vary inversely with the wind metallicity. For a given emission measure,
the mass of emitting gas follows directly from the volume of the emission
region and the volume filling factor, $f$, of \x emitting gas.
\fig~\ref{fig:mo} illustrates the inferred wind mass 
for various values of the alpha element abundance.  The points correspond to
the spectral models of Table~\ref{tab:alpfe}.\footnote{
	To convert emission measure to mass, we assume cylindrical volumes 
	for both the disk and halo components. 	The radius of each was 
	taken as 1~kpc.  The halo is assumed to extend to 1~kpc and the 
	disk to 0.20~kpc.  A filling factor of unity,$f \sim 1$, is shown.}
The spectral fitting constrains the alpha element abundance to be greater than 0.25
solar and the wind mass less than $6.2 \times 10^6$\msun.

The spectral fitting allows alpha element abundances higher than solar, and those 
solutions imply a lower mass of hot gas.   We use the mass of stellar ejecta
to place a generous lower limit on the wind mass.  From Table~\ref{tab:sb},
the mass of stellar ejecta is $M_{ej} = 9.8 \times 10^4$\msun\
(10 Myr) to $3.8 \times 10^5\msun$ (20 Myr).

For the best fitting model, $Z_{\alpha} = 1.0\zsun$, the inferred mass of 
the \x halo is $2.80 \times 10^6 f^{0.5}$\msun. The disk holds nearly as much 
hot gas, roughly $7.34 \times 10^5 f^{0.5}$\msun. The total mass of hot gas 
is then $3.53 \times 10^6 f^{0.5}$\msun. 
For the continuous star formation model, the mass loading,
$\chi \equiv (M_X - M_{ej}) / M_{ej}$,  is $\chi = 35$ to 8
for the age range 10 to 20~Myr.
Figure~\ref{fig:mo} shows how much the inferred wind
mass decreases for higher values of the alpha element abundance. For example,
if the wind contains equal masses of stellar ejecta and entrained gas,
i.e., $\chi = 1.0$, then $M_X = 7.6 \times 10^5$\msun, and the alpha elements
in the wind would be enriched to 40 times solar.  Numerical simulations
of the M82 superwind require a similar mass loading factor, $\sim 5$,
to reproduce the \x emission from that wind.
These mass loading factors are within the range suggested by
numerical simulations of winds (e.g., Suchkov \et 1996).




\subsubsection{Dynamical Evidence for a Gaseous Halo?}

In \S~\ref{sec:other} we suggested that the spatial correlation between
the hardest \x emission in the halo and the brightest regions of the halo
might indicate a halo shock. In the absence of a gaseous halo, the
wind would expand freely into the intergalactic medium, cooling adiabatically,
after breaking through the gaseous disk.  The high \x surface brightness
and absence of a radial temperature gradient are inconsistent with
this scenario. To generate a strong wind -- halo interaction, the halo 
must be important dynamically. This criterion is certainly met if
the swept-up mass of halo gas is comparable to the wind mass.  For
the minimum wind mass (which is 9 times smaller than our best estimate),
this condition requires a halo density 
$n_h \ge 2.5 \times 10^{-3} (M_{ej}/3.8 \times 10^5)$ \nden.  
The implied column density could have eluded direct detection in the
HI maps of  Roberts (1968) and Stil \& Israel (1998) but might be
reachable by the ultra-deep observations of M\"uhle \et (2001).


In a scenario where the wind generates a halo shock, the temperature of 
shocked halo gas would be directly related to the temperature of the wind 
emanating from the disk. Suppose we associate the hot, 
$0.7$~keV, thermal component with this wind generation region.  Then the wind 
emerges from the disk at roughly the sound speed, 430\kms, and  accelerates 
toward a terminal velocity $\sim \sqrt{3} c_s \sim 740$\kms\ at large
distances. The measured
temperature of the {\it green} regions, $0.3$~keV, corresponds to the 
temperature just behind a 550\kms adiabatic shock.  It appears at least
plausible then, that expansion of the hotter thermal component generates a 
shock which dominates the emission at 0.3~keV.  

The conundrum is that the shock front seen in the optical emission lines
indicates substantially lower expansion velocities.
The optical line ratios indicate most of the warm halo gas is  photoionized,
but the ratios do become more shock-like far out in the halo (Martin 1998)
and are consistent with shock speeds $\sim 100$\kms, consistent with
the expansion  velocity of the \Ha filaments (60 to 120\kms along the 
sightline). The expected postshock temperature, about $4.5 \times 10^5$~K, 
is too low to produce emission in the Chandra band.  Direct measurement
of the kinematics of the hot outflow would determine whether it has
significant bulk motion, and therefore whether the two shocks are distinct.  

The observational implications of the \x limb-brightening are quite 
significant. The interaction of the wind with this halo boosts the \x 
luminosity of the halo. The total luminosity of the 0.3~keV component is 
$5.0 \times 10^{38}$\ergsec, which is comparable to that in
the 0.7~keV component, $3.2 \times 10^{38}$\ergsec. 
 The main effect is simply to
make the wind brighter without requiring extreme mass loading.
 Since 
$L_x$ is still much less than the energy injection rate $L_W$, the radiative 
losses are far too small to stall the wind.  Hence we expect the wind to
blow through the halo and reach the IGM.

Comparison to the grid of models described by Silich and Tenorio-Tagle (2001)
indicates the wind is still strong enough to escape from the halo.
For the continuous star formation burst model, we find 
$<L_w> = 2.1 \times 10^{40}$\ergsec\ ($<L_w> = 8.3 \times 10^{40}$\ergsec)
at an age of 10~Myr (20~Myr).  
It is not clear whether this halo is typical of dwarf galaxies.
Tidal debris from the interaction with the 
NGC1569-HI cloud could be the source of some (most?) of the halo gas.

\subsection{Chemical Evolution}

One expects the stars which power the galactic
outflow to also enrich it.  Indeed, simulations suggest the wind removes the
new metals very efficiently (MacLow \& Ferrara 1999), but we can
now examine the efficiency directly.

\subsubsection{Wind Abundance and Mass Entrainment}

The wind carries metals recently ejected by supernovae and metals entrained 
from the ambient ISM. Figure~\ref{fig:chem} shows the resulting oxygen 
abundance of the wind for various mixing ratios of supernovae ejecta and
ISM.

The wind metallicity is the sum of metals from supernovae and the entrained
ISM divided by the wind mass, $M_x$. 
	\begin{equation}
	Z_{\alpha} = M_{ej} (Z_{\alpha,SN} + \chi Z_{\alpha,ISM}) / M_x.
	\end{equation}
Using the definition of the mass loading factor $\chi$,
	\begin{equation}
	M_x / M_{ej} = 1 + \chi,
	\end{equation}
to re-write Equation~1 in a form independent of mass, we have
	\begin{equation}
	Z_{\alpha} = (Z_{\alpha,SN} + \chi Z_{\alpha,ISM}) / (1 + \chi).
	\end{equation}
Similarly, the abundance ratio of alpha elements to iron in the wind
will be,
	\begin{equation}
	Z_{\alpha}/Z_{Fe} = (Z_{\alpha,SN} + Z_{\alpha,ISM}) /
	(Z_{Fe,SN} + Z_{Fe,ISM})
	\end{equation}
The curves in \fig~\ref{fig:chem} simply illustrate these relations for
various the degrees of mixing. The transition from ejecta-dominated 
metals to ISM-dominated metals shows up as an inflection in the curves.
The IMF-averaged metallicity of supernova ejecta is  $\sim 8$ times solar
for oxygen and $\sim 0.2$ times solar for Fe (model~A runs of Woosley \& 
Weaver (1995) at 0.1\zsun).  The oxygen and iron yields of massive stars 
are only known to an accuracy of a factor of two to three. We plot the
models for three plausible values of the supernova metallicity.

Each curve can be parameterized by the mass of stellar ejecta.
The spectral degeneracy between the wind metallicity
and the emission measure gives a unique mass for the \x emitting wind
at each point (recall \fig~\ref{fig:mo}).  At any point in \fig~\ref{fig:chem},
the mass of ejecta then follows directly from the wind mass and mass-loading 
factor. The actual value of $M_{ej}$ likely lies between the two values for
the continuous star formation scenario in Table~\ref{tab:sb}.  These points
are marked.  A starburst duration of 20~Myr gives the maximum mass of
ejecta and the lower limit on the mass loading factor.  The corresponding
upper limit on the wind metallicity is $O/H \sles\ 2 \osun$. Hence,
even though the spectral analysis cannot rule out a very high metallicity
wind, we find it would require an extremely low mass loading factor.  Given
the measured wind mass (for high metallicity), the low mass loading factors
require more supernova ejecta than the current burst is likely to have
produced.

We conclude that self-consistent chemical and dynamical models for
the wind require the mass loading factor to be $\sim 9$.  
Since the metallicity of the supernova ejecta is about $8 / 0.2 = 40$
times larger than the ambient ISM in \n1569, the metals from
the recent starburst dominate unless the mass loading factor exceeds
40. Most of the metals in the outflow come from the supernovae ejecta not 
the entrained ISM even though the ISM supplies the majority of the mass in 
the outflow. This result explains why the measured ratio of alpha elements 
to Fe is higher than solar. We regard the model predictions for the alpha
to iron abundance ratio in the bottom panel of \fig~\ref{fig:chem}
very tentatively owing to the large uncertainties in the relative yields.  
It is encouraging, however, that the mass loading factors required to 
match the measured $Z_{\alpha}/Z_{Fe}$ value, see the end of \S~4.2.2,
are so close to our best estimates.

%

\subsubsection{Metals Produced by the Burst vs. Metals in Wind}

The parameters of our best model for the integrated spectrum, 
$Z_{\alpha} = 1.0\zsun$ in Table~\ref{tab:alpfe}, imply the hot wind 
carries $3.5 \times 10^6$\msun\ of gas which includes $3.4 \times
10^4\msun$ of oxygen.  Models for higher abundance winds are less massive, 
so the mass of ejected metals is better constrained than the wind metallicity.
Inspection of \fig~\ref{fig:mo} suggests the mass of oxygen ejected in
the hot wind scales, roughly, as $M_O \approx 3.4 \times 10^4 \sqrt{Z_{\alpha}/
\zsun_{,\alpha}} \msun$.  The oxygen mass held by the warm phase ISM is considerably
lower. Although the total mass is similar, the metallicity is down by a factor 
$\sim 5$.  Our new measurement of the metallicity of the HI disk 
($\sim 0.20\zsun$) strongly suggests the neutral gas disk holds the majority 
of the gas phase oxygen, about $1.6 \times 10^5 (M_{HI}/8.4 \times 10^7)$\msun\ 
of oxygen.\footnote{
	Scaled to a distance d=2.2~Mpc, the HI measurements of
	Reakes imply $M_{HI} = 8.4 \times 10^7\msun$, which
	is considerably lower than the single dish measurement
	of Roberts ($1.6 \times 10^8\msun$).  However, 
	deep interferometric observations suggest an even larger
	value $M_{HI} \sim 2.4 \times 10^8\msun$ (M\"uhle \et 2001).
	We take the smaller measurement to be representative
	of the {\it disk} and associate the larger values with
	the very extended halo component.}
This disk enrichment shows that the metals were recycled rather than ejected
during some, presumably {\it quiescent}, periods of star formation.

It is interesting to compare the mass of oxygen carried by the wind to the
mass produced by the starburst. Table~\ref{tab:sb} shows the mass of oxygen 
produced for
two fiducial star formation scenarios.  The oxygen production from the 
Woosley \& Weaver (1995) models was integrated over the stellar IMF. The 
least massive Type~II 
supernovae progenitors live $\sim 40$~Myr and have probably not yet exploded.  
Comparison of the 5th and 6th lines in the Table shows the correction for
this time lag is small relative to variation in the yield among the 
supernova calculations (e.g., Gibson \et and references therein).
The instantaneous burst model for the super star clusters produces
only about one-tenth of the actual mass of oxygen in the outflow.
The yield of nearly the full starburst population is apparently required
to explain the metal content of the wind.  This result is consistent
with the previous conclusion that the super star clusters alone cannot
power the wind.

\subsubsection{Retention of Metals and Burst Duty Cycle}

Nearly all the oxygen produced by 
the current burst is required to enrich the wind, so a previous generation of 
stars must have produced the metals held by the HI disk. The galaxy presumably
retained metals during periods when no galactic wind was generated.  Hence,
we refer to these earlier epochs, when the star formation rate was lower and
metals were recycled, as periods of quiescent star formation.  If we adopt $M_{HI}
= 8.4 \times 10^7$\msun\ as the mass of disk gas that is enriched to 
$\sim 0.20$\zsun, then the mass of oxygen in the wind is about one-fifth of 
that held by the HI disk.  Hence, the mass of stars that formed in the 
quiescent mode must be five to ten times larger than the stellar mass formed 
by the current 
burst, where the larger value applies for $M_{HI} = 1.6 \times 10^8$\msun.
Extending the stellar IMF down to 0.1\msun, the mass of stars formed in
the quiescent phase is  $M_* \approx 4 - 8 \times 10^7 (\tau/20~Myr)\msun$.

The mass of stars required to enrich the HI disk can also be estimated
from the metallicity of the disk and the total gas content of the system.
If little mass was ejected from the galaxy previously, then the amount of 
quiescent star formation required to enrich the HI disk to a metal fraction 
$Z$ is
	\begin{equation}
	M_*(t) = (e^{Z/y} -1) M_g(t),
	\end{equation}
assuming instantaneous recycling of the metals.  For an oxygen abundance of
20\% solar, the oxygen mass fraction is $Z = 0.0019$.  Substituting the oxygen 
yield $y = 0.0061$, which is appropriate for a lower mass cutoff 0.1\msun\  
(Woosley \& Weaver 1995), into Equation~4 yields $M_* = 0.36 M_g(t),$ 
or $M_* \approx 3 - 6 \times 10^7$\msun.  Since these values are similar to
our first estimate, it seems unlikely that large amounts of gas have already 
been expelled from \n1569.  However, both estimates are strickly lower limits
on the stellar mass.  The difference of  the dynamical mass of the disk 
$3.3 \times 10^8$\msun\ (Israel 1988) and the gas mass yields an upper limit
on the stellar mass of $1.7 - 2.5 \times 10^8$\msun.

Uncertainty about the initial gas mass of the galaxy and the current stellar 
mass 
prevents us from drawing definitive conclusions about the fraction of total
metals ejected from the galaxy by starburst winds.  For purposes of illustration,
however, Figure~\ref{fig:chemev3} shows a simple model of the enrichment history.
For a closed box model to be consistent with the observed ISM abundance, the
stellar mass must be near our minimum values.  It is quite possible, however, that
the stellar mass and gas mass are roughly equal. In this case,
\fig~\ref{fig:chemev3} demonstrates that 30\% to 50\% of the initial gas mass 
could have been blown away (or stripped from the galaxy).




\section{Summary}

Galactic winds influence the chemical evolution of galaxies and enrich the 
intergalactic medium.  Establishing the efficiency of mass and metal ejection
has required empirical calibration, however, due to the complexity
of exchange processes between phases of the ISM and the sensitivity of 
blowout to the structure of the gas disk.  The result that might not
have been anticipated is that, when a galaxy undergoes a starburst,
most of the metals expelled by Type~II supernovae are ejected from the
galaxy even though the bulk of the ISM remains intact.  The metals remain
in the hot phase on a timescale that is long compared to the blowout
timescale of the wind.  The abundance of alpha elements is high relative 
to iron which is consistent with a Type~II supernova-driven wind. 

Our main results are as follows.

\begin{enumerate}

\item{
	We resolve the hard emission into a population of point sources.
The resolved sources contribute 57\% ( 77\%) of the 1.1 -- 6~keV (2 --6~kev)
emission, and the correction for unresolved point sources is estimated to be
no more than 18\% of the flux from the resolved population.
The majority of the sources are believed to be high-mass \x binaries due to
their variability, their spectral hardness, and the paucity of compact
radio sources in \n1569. A new \x bright supernova remnant, 
CXOU043054.1+645043, is reported in the disk of \n1569.  The low number of SNRs
with detectable radio or \x emission in \n1569 is consistent with at least
half of the supernovae exploding in low density environments where they
radiate inefficiently -- consistent with the generation of a galactic wind
(e.g., Chevalier \& Fransson 2001).  
	}

\item{
	In contrast to previous analyses (Della Ceca \et 1996),  we
find significant \x absorption from the disk of \n1569.  The resulting color
gradient is seen unambiguously in the spectral imaging.  Including it
in the spectral modeling allows a higher continuum normalization and
stronger emission lines.  Requiring the fitted photoelectric absorption from the
disk to be consistent with the HI column implied by the 21-cm brightness 
temperature contrains the metal abundance of the ambient gas disk to be
greater than 0.1\zsun\ and less than 1\zsun\ which allows a value equal
to that of the HII regions.  We conclude that the neutral disk holds
a higher metal mass than the wind or HII regions.
	}

\item{We confirm the strong spatial correlation between the extended \x 
emission and the \Ha filaments and discover that the halo consists of
two  \x emission components.  The brighter, 0.3~keV, component is associated
with the halo shock generated by the outflow. This emission comes from the
highest \x surface brightness regions of the halo which are found near the
\Ha filaments. The \x emission is detected over the full extent of the 
\Ha nebula but not convincingly beyond these filaments.  The \x emission
may come from the mixing layers between the shock and the bubble interior
rather than the actual shock front.  In either scenario, however, the presence
of the shock implies the wind encounters a gaseous halo that was previously 
unrecognized. The wind -- halo interaction significantly boosts the \x 
luminosity of the galaxy.  We argue that the wind in \n1569 is likely powerful
enough to blow through this halo and escape the galactic potential.
The faint emission component in the halo is likely the wind fluid itself
and may fill much of the volume.  }

\item{The wind metallicity must be greater than 0.25\zsun\ in order
to fit the strong Mg and O lines in the spectrum.  The large
residuals in these lines in otherwise degenerate models with low
metal abundance could not be recognized in lower signal-to-noise data.
The spectrum alone provides no constraint on the maximum metallicity
of the wind owing to a degeneracy between models with continuum
normalization and  metallicity.    We can already conclude, however,
that the wind metallicity does not exceed $O/H \approx 2 \osun$.  Otherwise,
the continuum normalization is pushed below that generated by the
starburst ejecta alone which is known to within a factor of a few
from simple population synthesis.  Such metals predict that the
burst produced 34,000 to 44,000\msun\ of oxygen in 10 -- 20 Myr.  The wind 
carries about 34,000\msun\ of oxygen. Hence, we conclude that the hot wind 
transports most of the metals synthesized by the burst.}

\item{
	We find the ratio of alpha elements to Fe is higher than
the solar ratio. Supernovae yields are only known to within a factor
of few, but the measured alpha element to iron ratio is clearly higher
than the solar ratio and is consistent with a significant alpha element
contribution from Type~II supernova ejecta.  This result is at first 
surprising given that most of the mass in the wind is entrained from the 
surrounding ISM. (We estimate $\chi \sim 9$).  However, since \n1569 is a 
dwarf galaxy, the mean metallicity of the ISM is only 0.20\zsun, and the mass 
loading would have to be more than a factor of 40 before the entrained oxygen
mass exceeded the oxygen mass in the SNe ejecta. 
	}

\item{
	The mass of oxygen leaving the galaxy in the wind is about 10\% to 17\% of
the oxygen that was recycled into the neutral gas disk.  Periods with lower star 
formation rates, i.e., quiescent star formation, must have produced at least five 
times as many stars (and metals) as the current burst.  The total stellar mass
and the initial gas mass are poorly contrained, so our chemical evolution models
allow quite a large range for the ejected (or stripped) mass fractions 
over cosmic time.
	}

Our results demonstrate that starbursts in dwarf galaxies do pollute
the intergalactic medium.   The  question of whether dwarf or giant galaxies 
dominate this enrichment remains open (De Young \& Gallagher 1990).  However, 
the enrichment from
dwarfs is particularly interesting in light of the large numbers of dwarf 
galaxies expected in the early universe, the measured enrichment of the 
Lyman alpha forest at redshift 3 to 4 (e.g., Songailia \& Cowie 1996), and
the low metallicities of many nearby dwarfs (Vader 1986).
A definitive measurement of a supersolar alpha-element to iron abundance ratio 
in the IGM would allow one to conclude that starburst winds,  rather
than winds generated during non-starburst periods (e.g., Efstathiou 2000) or 
tidal debris (e.g., Gnedin 1998), dominate the IGM enrichment.




\end{enumerate}

\acknowledgements

We thank Deidre Hunter for the use of her HST F555 image and Eric
Wilcots and Stephanie M\"uhle for previews of their VLA 21-cm neutral 
hydrogen maps.  Cornelia Lang reduced the 20 cm VLA data, and Nicole Homeier
assisted with the optical observations.  We thank Kim Weaver for providing
comments on a draft of the paper.
This research made use of the NASA/IPAC Extragalactic Database 
(NED) which is operated by the Jet Propulsion Laboratory, California
Institute of Technology, under contract with the National Aeronautics and 
Space Administration. C.L.M. thanks the Sherman Fairchild Foundation for
financial support.  This work could not have been completed without funding 
from Smithsonian Astrophysical Observatory Award No. GO0-1140A under
NASA contract No. NAS8-39073.


\clearpage
	\input table1.tex
	\input table2.tex


\begin{table}
\tiny
\caption{Models Fitted to Integrated Spectrum} \label{tab:int}
\begin{tabular}{llllllll}
\hline
model             		& \nh  	&kT     &norm    	& \nh	&kT, Gamma &norm    &$\chi^2_{\nu}  ~~\nu$ \\
				& (\col)  & (keV)   & (\norm)		& (\col)  & (keV, \#)  & (\norm,  keV$^{-1}$cm$^{-2}$) & \\
\hline
\hline
wabs(vabs*M)		& $4.1 \times 10^{21}$ & 0.62 & $1.19 \times 10^{-3}$ & & & & 3.12 160 \\
DGHM96 wabs(M+M) 		&0	&[0.63] &$4.29 \times 10^{-4} $	&0	&[3.81]   &$3.30 \times 10^{-4} $&1.83 161	\\
wabs(M+M) 		&0 	&0.66   &$4.53 \times 10^{-4} $	&0	&2.70     &$2.93 \times 10^{-4} $&1.72 159	\\
DGHM96 wabs(M+P)    		&0 	&[0.64] &$3.66 \times 10^{-4} $	&0	&[2.09]   &$1.13 \times 10^{-4} $&2.14 161 \\
wabs(M+P)    		&0 	&0.67   &$4.65 \times 10^{-4} $	&0	&1.73     &$0.72 \times 10^{-4} $&1.79 159 \\
wabs(vabs*M+vabs*M)	    & $2.5 \times 10^{21}$ &0.27   &$1.23 \times 10^{-3} $ &$1.1 \times 10^{22}$ &0.73  &$1.41 \times 10^{-3} $&1.21 157 \\
wabs(vabs*Pt+vabs*M+vabs*M) & $3.9 \times 10^{21}$ &0.25   &$1.55 \times 10^{-3} $ &$6.0 \times 10^{21}$ & 0.73 &$6.23 \times 10^{-4} $&1.20 157 \\
\hline
\end{tabular}
Notes -- 
The abundances are fixed at 0.25\zsun. The $[...]$ notation denotes a fixed parameter.
The foreground absorption component has solar metallicity and a fixed column
of 2.1e21\col. The models are MEKAL (M), power law (P), photoelectric absorption with solar abundance (wabs), 
and photoelectric absorption with 0.25\zsun abundance (vabs). The power law fitted to the combined
point sources (Pt) has $\Gamma = 2.40$, and $norm = 8.40 \times 10^{-5}$~photons keV$^{-1}$~cm$^{-2}$~s$^{-1}$~(at 1~kev) 
with an intrinsic (0.25\zsun) absorbing component of $N_H = 2.1 \times 10^{21}\col$. 

\normalsize
\end{table}
\begin{table}
\tiny
\caption{Models Fitted to S1-5 Spectrum} \label{tab:s1-5}
\begin{tabular}{lllllllllll}
\hline
model					&NH	&kT	&norm	&kT	&norm	&NH	&kT	&norm	&chi2v	&v 	\\
					& (\col)& (keV) &(\norm)&(keV)  &(\norm)&(\col) & (keV) &(\col) &       &       \\
\hline
\hline
wabs(vabs*M+vabs*M)$_{T1}$	&$2.0\times 10^{17}$	&[0.61]	&$1.8\times 10^{22}$	& 0	& 0	 & $1.8 \times 10^{22}$	&[0.61]	&$2.77 \times 10^{-4}$	&1.57	&129	\\
wabs(vabs*M+vabs*M)		&$1.9\times 10^{21}$	&0.29	&$3.46\times 10^{-4}$	&0	&0	 & $6.7 \times 10^{21}$	&0.73	&$2.63 \times 10^{-4}$	&1.22	&128	\\	
wabs(vabs*M+M+vabs*M)	&$3.9\times 10^{21}$	&0.28	&$6.70\times 10^{-4}$	&0.1	&$7.48\times 10^{-4}$ &$8.22\times 10^{21}$	&0.73	&$2.59\times 10^{-4}$	&1.22	&126	\\
wabs(vabs*M+M+vabs*M)	&$8.1\times 10^{22}$    & 0.10	&41.69			&0.34	&$1.47\times 10^{-4}$ &$3.0\times 10^{21}$	&0.74	&$1.81\times 10^{-4}$	&1.17	&127	\\
\hline
\end{tabular}
Notes -- The components are as described in the notes to Table~3.
Point sources have been masked out, so their power law component is not included.

\normalsize
\end{table}
\begin{table}[h]
\tiny
\caption{Variable Abundance Models Fitted to Integrated Spectrum} \label{tab:alpfe}
\begin{tabular}{lllllllll}
\hline
$Z_{\alpha}$	& $Z_{Fe}$	& $\chi_{\nu}^2$	& $N_{H,1}$	& $kT_1$& $norm_1$		& $N_{H,2}$	& $kT_2$	& $norm_2$	\\
(\zsun)		& (\zsun)	&			& (\col)	& (keV)	& (\norm)		& (\col)	& (keV)		& (\norm) \\
\hline
\hline
0.10		& 0.12		& 1.364		& $4.38 \times 10^{21}$	& 0.22	& $1.00 \times 10^{-2}$	& $1.80 \times 10^{21}$	& 0.72	& $6.44 \times 10^{-4}$\\
0.25		& 0.14		& 1.246		& $2.98 \times 10^{21}$	& 0.31	& $2.19	\times 10^{-3}$	& $1.60 \times 10^{21}$	& 0.71	& $4.23	\times 10^{-4}$	\\
0.50		& 0.21		& 1.195		& $3.09 \times 10^{21}$	& 0.31	& $1.36	\times 10^{-3}$	& $1.43 \times 10^{21}$	& 0.70	& $2.76	\times 10^{-4}$	\\
1.00		& 0.37		& 1.185		& $3.0  \times 10^{21}$	& 0.31	& $6.40	\times 10^{-4}$	& $1.5  \times 10^{21}$	& 0.65	& $2.19	\times 10^{-4}$	\\
5.00		& 1.41		& 1.194		& $5.5  \times 10^{21}$	& 0.25	& $2.81 \times 10^{-4}$	& $4.1  \times 10^{20}$	& 0.61	& $5.90	\times 10^{-5}$	\\
10.0		& 4.26		& 1.200		& $3.2  \times 10^{21}$	& 0.31	& $9.51 \times 10^{-5}$	& $3.8  \times 10^{21}$	& 0.70	& $2.58	\times 10^{-5}$	\\
100		& 43		& 1.197		& $3.8  \times 10^{21}$	& 0.25	& $1.10 \times 10^{-5}$	& $2.6  \times 10^{21}$	& 0.61	& $3.89	\times 10^{-6}$	\\
\hline
\end{tabular}

Notes -- All models contain a fixed Galactic absorption component ($1.2 \times 10^{21}\col$, 1.0 \zsun)
and  power law describing the point source component ($\Gamma = 2.40$, and $norm = 8.40 \times 10^{-5}$~photons 
keV$^{-1}$~cm$^{-2}$~s$^{-1}$~at 1~kev) with an intrinsic (0.25\zsun) absorbing component of $N_H = 2.3 
\times 10^{21}\col$. The alpha element abundance of the two thermal components is fixed at the 7 levels shown.
The other elements are varied with iron as the temperatures and normalizations of the two hermal components
are fitted.  The instrinsic absorbing columns for each thermal component are varied independently and were
assumed to solar abundance in this exercise.

\normalsize

\end{table}


\begin{table}
\caption{Power and Stellar Ejecta Supplied by Massive Stars} \label{tab:sb}
\begin{tabular}{lllll}
\hline
SFH\tablenotemark{a}	& C-SFR	& C-SFR & I-Burst	& I-Burst \\
Age	& 10~Myr& 20~Myr & 10~Myr	& 20~Myr \\
\hline
\hline
SFR\tablenotemark{a} (\msunyr) 		& 0.17	& 	0.17	& \nodata	& \nodata \\
$M_{*}$\tablenotemark{b} (1-100\msun)	& $1.7 \times 10^6$	& $3.4 \times 10^6$	& $3.1 \times 10^5$	&$3.1 \times 10^5$\\
$M_{ejecta}$\tablenotemark{b} (\msun) 	& $9.8 \times 10^4$	& $3.8 \times 10^5$	& $4.5 \times 10^4$	&$8.7 \times 10^4$\\
$E_{W}$\tablenotemark{b} (ergs) 		& $6.8 \times 10^{54}$	& $2.6 \times 10^{55}$	& $3.1 \times 10^{54}$	&$6.9 \times 10^{54}$\\
$M_O (t)$\tablenotemark{c} (\msun)& \nodata & \nodata & $3.4 \times 10^3$ & $4.4 \times 10^3$ \\
$M_O$(final)\tablenotemark{c} (\msun) & $2.5 \times 10^4$ & $5.1 \times 10^4$ & $4.8 \times 10^3$ & $4.8 \times 10^3$ \\
\hline
\end{tabular}

\tablenotetext{a}{
	This table illustrates the mechanical, radiative, and 
	chemical feedback in \n1569 for two star formation histories -- an
	instantaneous burst (I-Burst) and a constant star formation rate (C-SFR).  
	To illustrate the relative contribution of the two prominent super star clusters
	to the more widespread massive star population, we normalize the C-SFR
	model the the former using the total \Ha luminosity of the galaxy (Martin \&
	Kennicutt 1997), but we normalize the I-Burst to the stellar mass of the two SSCs
	(Ho \& Filippenko 1996).
	}
\tablenotetext{b}{
	Based on population synthesis models from Leitherer \et (1999) which
	use a Salpeter IMF from 1 to 100\msun and solar abundances.  (Using
	0.2\zsun models makes negligible differences for our results.)  The 
	lifetime of the lowest mass Type~II supernova progenitors is is 40~Myr 
	in this model, so  the instantaneous burst models reach their final yield, 
	and the constant star formation rate model has not attained equilibrium rates.
	The value shown is the integral of the rate over 10 or 20 Myr as indicated.
	}
\tablenotetext{c}{
	We integrate the oxygen yields (Woosely \& Weaver 1995) over the stellar intial 
	mass function. We assume the least massive star  that produces a Type~II 
	supernovae has an initial stellar mass of 10\msun.  Stars with initial masses 
	larger than about 30\msun\ have considerable  reimplosion of heavy elements, and 
	no star more massive than 40\msun\  contributes to the yield in this model.
	}  


\end{table}

\clearpage

\input fig_cap.tex

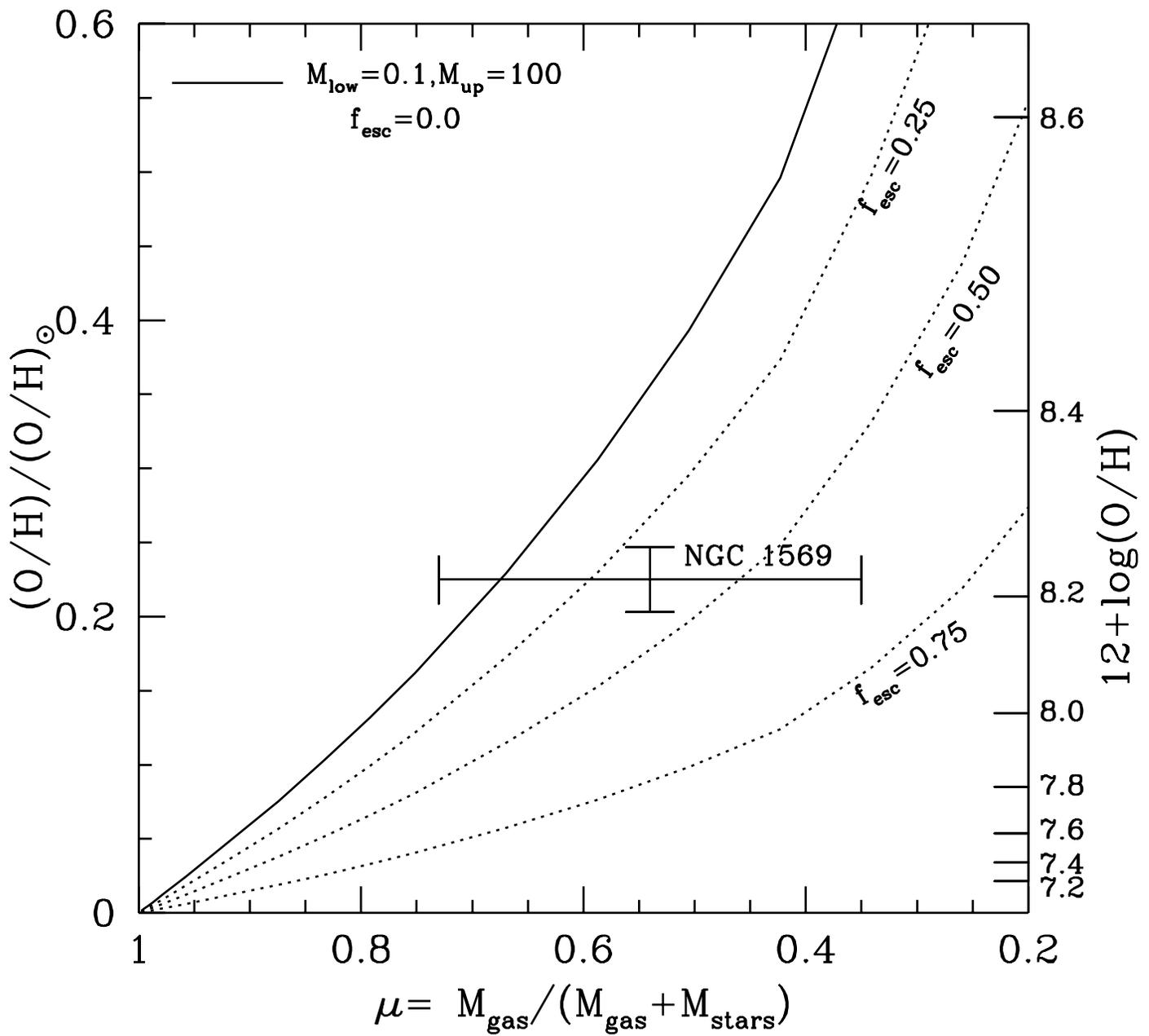
\end{document}

%% file: table1.tex
\begin{deluxetable}{rrrrrrrrrrrrrrrrr}
\label{srctable.tab}
\rotate
\tabletypesize{\scriptsize} 
\setlength{\tabcolsep}{0.02in} 
\tablewidth{8.7in}
\tablecaption{NGC 1569 X-ray Point Sources}
\tablehead{
\colhead{ID} & 
\colhead{CXO Designation} & 
\colhead{RA 2000} & 
\colhead{Dec 2000} & 
\colhead{$C_{tot}$} & 
\colhead{$C_{S}$} & 
\colhead{$C_{M}$} & 
\colhead{$C_{H}$} & 
\colhead{$R_{PSF}$} & 
\colhead{$S_\nu(3)$ } & 
\colhead{err} & 
\colhead{$S_\nu(20)$} & 
\colhead{err} & 
\colhead{R~mag} & 
\colhead{err} & 
\colhead{$\log(f_X/f_V)$} & 
\colhead{Type}   \\ 
\colhead{(1)} & 
\colhead{(2)} & 
\colhead{(3)} & 
\colhead{(4)} & 
\colhead{(5)} & 
\colhead{(6)} & 
\colhead{(7)} & 
\colhead{(8)} & 
\colhead{(9)} & 
\colhead{(10)} & 
\colhead{(11)} & 
\colhead{(12)} & 
\colhead{(13)} & 
\colhead{(14)} & 
\colhead{(15)} & 
\colhead{(16)} & 
\colhead{(17)}  }
\startdata
 1 & CXOU043023.0+644913  & 04 30 23.05 & +64 49 13.9 &    16.4$\pm$5.0 &     1.5$\pm$1.6 &    1.9$\pm$1.9 &      15$\pm$4.7 & 0.97 &  $<$0.09 & \nodata &  $<$0.12 & \nodata & $>$23.7 & \nodata & $>$1.27 & AGN    \\
 2 & CXOU043025.2+645035  & 04 30 25.22 & +64 50 35.6 &    81.2$\pm$9.5 &     2.5$\pm$1.9 &   11.1$\pm$3.8 &    68.5$\pm$8.8 & 0.91 &  $<$0.09 & \nodata &  $<$0.12 & \nodata & $>$23.7 & \nodata & $>$2.87 & AGN    \\
 3 & CXOU043027.1+644926  & 04 30 27.15 & +64 49 26.1 &    16.9$\pm$5.2 &     2.1$\pm$1.9 &    4.1$\pm$2.8 &    11.5$\pm$4.3 & 1.39 &  $<$0.09 & \nodata &  $<$0.12 & \nodata &   17.39 &   0.01 &   -4.47 & Star   \\
 4 & CXOU043030.9+645205  & 04 30 30.91 & +64 52 05.3 &  182.8$\pm$14.0 &    10.6$\pm$3.8 &   39.9$\pm$6.8 &    133.2$\pm$12 & 0.77 &  $<$0.09 & \nodata &    0.15 &    0.03 &   21.28 &   0.07 &    1.46 & AGN    \\
 5 & CXOU043032.8+644735  & 04 30 32.83 & +64 47 35.3 &    27.3$\pm$6.0 &     1.5$\pm$0.8 &    3.1$\pm$0.7 &      28.1$\pm$6 & 1.05 &  $<$0.09 & \nodata &  $<$0.12 & \nodata & $>$23.7 & \nodata & $>$1.78 & AGN    \\
 6 & CXOU043034.3+644740  & 04 30 34.31 & +64 47 40.4 &  392.5$\pm$20.4 &    21.9$\pm$5.2 &   79.2$\pm$9.3 &  292.4$\pm$17.7 & 1.01 &  $<$0.09 & \nodata &  $<$0.12 & \nodata &   21.42 &   0.08 &    2.35 & AGN    \\
 7 & CXOU043035.2+645222  & 04 30 35.22 & +64 52 22.1 &    14.3$\pm$4.4 &     1.5$\pm$0.7 &    1.5$\pm$1.5 &    13.5$\pm$4.2 & 1.11 &  $<$0.09 & \nodata &  $<$0.12 & \nodata & $>$23.7 & \nodata & $>$1.13 & AGN    \\
 8 & CXOU043036.5+644921  & 04 30 36.53 & +64 49 21.4 &    13.2$\pm$4.1 &     0.5$\pm$0.5 &    2.2$\pm$1.9 &    11.5$\pm$3.8 & 1.42 &  $<$0.09 & \nodata &  $<$0.12 & \nodata & $>$23.7 & \nodata & $>$1.05 & AGN    \\
 9 & CXOU043039.0+645012  & 04 30 39.03 & +64 50 12.8 &    41.4$\pm$7.1 &    19.5$\pm$4.8 &   17.4$\pm$4.7 &     5.4$\pm$3.0 & 1.36 &  $<$0.09 & \nodata &  $<$0.12 & \nodata &   17.11 &    0.07 &   -3.84 & Star   \\
10 & CXOU043039.5+645108  & 04 30 39.59 & +64 51 08.5 &    14.4$\pm$4.4 &     1.5$\pm$1.5 &    0.9$\pm$1.5 &    12.9$\pm$4.1 & 1.29 &  $<$0.09 & \nodata &  $<$0.12 & \nodata &    21.4 &    0.13 &   -0.96 & Star   \\
11 & CXOU043043.7+644943  & 04 30 43.79 & +64 49 43.0 &      12$\pm$3.9 &     2.2$\pm$1.9 &    4.2$\pm$2.5 &     6.5$\pm$2.9 & 1.16 &  $<$0.09 & \nodata &  $<$0.12 & \nodata &   19.05 &    0.02 &    -3.3 & Star   \\
12 & CXOU043044.2+644924  & 04 30 44.21 & +64 49 24.9 &     9.3$\pm$3.7 &     1.2$\pm$1.5 &    7.2$\pm$3.1 &     1.8$\pm$1.9 & 1.22 &  $<$0.09 & \nodata &  $<$0.12 & \nodata &   16.62 &    0.06 &   -5.78 & Star   \\
13 & CXOU043044.8+645247  & 04 30 44.86 & +64 52 47.0 &    40.5$\pm$6.8 &    11.5$\pm$3.8 &   14.5$\pm$4.2 &    15.5$\pm$4.3 & 0.82 &  $<$0.09 & \nodata &  $<$0.12 & \nodata &   18.94 &    0.02 &   -2.18 & Star   \\
14 & CXOU043046.9+645106  & 04 30 46.90 & +64 51 06.8 &    27.5$\pm$7.8 &       0.7$\pm$2 &    8.2$\pm$5.0 &    19.5$\pm$6.1 & 1.94 &  $<$0.09 & \nodata &  $<$0.12 & \nodata &    19.7 &     0.6 &   -1.87 & XRB    \\
15 & CXOU043047.5+645055  & 04 30 47.55 & +64 50 55.6 &    34.0$\pm$8.9 &     1.8$\pm$2.4 &    8.8$\pm$5.6 &    26.4$\pm$7.6 & 1.16 &  $<$0.09 & \nodata &  $<$0.12 & \nodata & $>$16.8 &     0.6 & $>$-4.3 & XRB    \\
16 & CXOU043048.1+645050  & 04 30 48.14 & +64 50 50.5 &   1150.5$\pm$35 &    139$\pm$12.5 & 462.4$\pm$22.4 &  550.1$\pm$24.3 & 1.16 &  $<$0.09 & \nodata &  $<$0.12 & \nodata & $>$16.8 &     0.6 & $>$-0.7 & XRB    \\
17 & CXOU043048.2+645046  & 04 30 48.25 & +64 50 46.6 &  159.3$\pm$13.6 &     3.5$\pm$2.5 &   19.8$\pm$5.3 &  136.9$\pm$12.5 & 1.30 &  $<$0.09 & \nodata &  $<$0.12 & \nodata & $>$16.8 &     0.6 & $>$-2.7 & XRB    \\
18 & CXOU043048.5+645049  & 04 30 48.57 & +64 50 49.6 &     9.7$\pm$6.9 &    19.4$\pm$5.1 &    2.6$\pm$3.9 &     4.1$\pm$3.3 & 1.33 &  $<$0.09 & \nodata &  $<$0.12 & \nodata & $>$16.8 &     0.6 & $>$-5.5 & XRB    \\
19 & CXOU043048.6+645058  & 04 30 48.61 & +64 50 58.5 &  637.2$\pm$26.2 &    32.1$\pm$6.4 & 148.6$\pm$13.1 &  457.5$\pm$22.1 & 1.52 &  $<$0.09 & \nodata &  \nodata & \nodata & $>$19.3 & \nodata &  $>$0.9 & Cluster \\
20 & CXOU043048.9+644940  & 04 30 48.93 & +64 49 40.1 &    10.9$\pm$3.8 &     2.3$\pm$1.9 &    8.3$\pm$3.3 &     1.1$\pm$1.5 & 1.27 &  $<$0.09 & \nodata &  \nodata & \nodata &    14.5 &    0.02 &   -7.56 & Star   \\
21 & CXOU043049.5+645042  & 04 30 49.56 & +64 50 42.6 &     1.9$\pm$4.1 &       0.8$\pm$2 &    2.9$\pm$2.6 &     4.0$\pm$3.0 & 1.75 &  $<$0.09 & \nodata &  \nodata & \nodata & $>$16.8 &     0.6 & $>$-7.2 & XRB    \\
22 & CXOU043049.8+645055  & 04 30 49.85 & +64 50 55.4 &  144.5$\pm$12.9 &     8.5$\pm$3.5 &   33.5$\pm$6.6 &  103.4$\pm$10.8 & 1.18 &  $<$0.09 & \nodata &  \nodata & \nodata & $>$19.9 & \nodata & $>$-0.0 & Cluster \\
23 & CXOU043050.2+645056  & 04 30 50.23 & +64 50 56.2 &    -4.6$\pm$4.9 &     1.6$\pm$1.6 &    1.8$\pm$3.1 &     6.2$\pm$3.8 & 1.29 &  $<$0.09 & \nodata &  \nodata & \nodata & $>$16.8 &     0.6 & $>$31.9 & XRB    \\
24 & CXOU043051.3+645050  & 04 30 51.38 & +64 50 50.9 &     5.9$\pm$4.9 &     2.1$\pm$1.9 &      2.2$\pm$3 &     4.4$\pm$3.7 & 1.56 &  $<$0.09 & \nodata &  \nodata & \nodata & $>$16.8 &     0.6 & $>$-6.0 & XRB    \\
25 & CXOU043051.7+645101  & 04 30 51.70 & +64 51 01.4 &     6.7$\pm$3.5 &     0.5$\pm$0.5 &    2.4$\pm$2.2 &       4.8$\pm$3 & 1.17 &  $<$0.09 & \nodata &  \nodata & \nodata & $>$16.8 &     0.6 & $>$-5.9 & XRB    \\
26 & CXOU043052.3+645042  & 04 30 52.34 & +64 50 42.3 &    11.9$\pm$4.7 &     2.9$\pm$2.2 &    1.4$\pm$2.6 &     8.4$\pm$3.7 & 1.54 &  $<$0.09 & \nodata &  \nodata & \nodata & $>$16.8 &     0.6 & $>$-5.3 & XRB    \\
27 & CXOU043053.3+645044  & 04 30 53.39 & +64 50 44.1 &    23.8$\pm$5.6 &     1.7$\pm$0.8 &    8.3$\pm$3.5 &    16.5$\pm$4.6 & 1.47 &  $<$0.09 & \nodata &  \nodata & \nodata & $>$16.8 &     0.6 & $>$-4.6 & XRB    \\
28 & CXOU043054.1+645043  & 04 30 54.11 & +64 50 43.1 &    24.9$\pm$5.7 &     1.5$\pm$0.8 &   12.9$\pm$4.2 &    13.0$\pm$4.1 & 1.53 &     0.69 &     0.1 &      5.6 &    0.10 & $>$16.8 &     0.6 & $>$-4.6 & SNR?   \\
29 & CXOU043057.4+645048  & 04 30 57.41 & +64 50 48.6 & 231.0$\pm$15.9 &     3.5$\pm$2.2 &   28.6$\pm$5.9 &  199.9$\pm$14.8 & 1.40 &  $<$0.09 & \nodata &  $<$0.12 & \nodata & $>$16.8 &     0.6 & $>$-2.4 & XRB    \\
30 & CXOU043058.0+644910  & 04 30 58.04 & +64 49 10.9 &    24.3$\pm$5.5 &     1.5$\pm$1.5 &    6.1$\pm$2.9 &    17.7$\pm$4.7 & 1.46 &     0.95 &    0.03 &     0.85 &    0.04 & $>$23.7 & \nodata & $>$1.66 & AGN    \\
31 & CXOU043058.3+644852  & 04 30 58.37 & +64 48 52.3 &      55.6$\pm$8 &    10.8$\pm$3.8 &   26.6$\pm$5.7 &    19.0$\pm$4.8 & 1.47 &  $<$0.09 & \nodata &  $<$0.12 & \nodata &   19.54 &    0.08 &   -1.31 & Star   \\
32 & CXOU043101.9+644818  & 04 31 01.92 & +64 48 18.4 &    25.4$\pm$5.6 &     1.5$\pm$0.7 &    6.5$\pm$2.9 &    19.6$\pm$4.9 & 1.52 &  $<$0.09 & \nodata &  $<$0.12 & \nodata &   22.01 &    0.13 &    0.16 & AGN    \\
33 & CXOU043105.1+645058  & 04 31 05.16 & +64 50 58.2 &     9.2$\pm$3.8 &     1.5$\pm$0.9 &    1.0$\pm$1.5 &     9.7$\pm$3.6 & 1.90 &  $<$0.09 & \nodata &  $<$0.12 & \nodata &   21.43 &    0.02 &   -1.38 & Star   \\
34 & CXOU043111.5+644947  & 04 31 11.52 & +64 49 47.1 &    24.3$\pm$5.5 &     0.5$\pm$0.5 &    7.3$\pm$3.1 &    17.5$\pm$4.7 & 1.32 &  $<$0.09 & \nodata &  $<$0.12 & \nodata &   18.77 &    0.02 &   -2.85 & Star   \\
35 & CXOU043113.2+645229  & 04 31 13.20 & +64 52 29.5 &  402.1$\pm$20.6 &      31.2$\pm$6 &  98.5$\pm$10.4 &  273.3$\pm$17.0 & 0.79 &     0.28 &    0.04 &     0.70 &    0.04 &    21.6 &    0.10 &    2.54 & Star   \\
36 & CXOU043113.5+645217  & 04 31 13.51 & +64 52 17.3 &    35.9$\pm$6.6 &     1.5$\pm$1.5 &    3.5$\pm$2.2 &    31.9$\pm$6.2 & 1.01 &  $<$0.09 & \nodata &  $<$0.12 & \nodata & $>$23.7 & \nodata & $>$2.05 & AGN    \\
37 & CXOU043114.0+645107  & 04 31 14.02 & +64 51 07.9 &  773.0$\pm$28.4 &  286.4$\pm$17.4 & 343.6$\pm$19.1 &  143.9$\pm$12.5 & 1.00 &  $<$0.09 & \nodata &  $<$0.12 & \nodata &   12.83 &    0.01 &   -4.83 & Star   \\
38 & CXOU043114.5+645024  & 04 31 14.54 & +64 50 24.5 &    57.3$\pm$8.1 &     1.3$\pm$1.5 &    5.5$\pm$2.7 &    51.5$\pm$7.7 & 1.13 &  $<$0.09 & \nodata &  $<$0.12 & \nodata & $>$23.7 & \nodata & $>$2.52 & AGN    \\
39 & CXOU043116.8+644950  & 04 31 16.85 & +64 49 50.0 & 4497.5$\pm$67.8 &  827.0$\pm$29.3 & 2160.5$\pm$47.0 & 1511.5$\pm$39.5 & 1.04 &  $<$0.09 & \nodata &  $<$0.12 & \nodata &   12.77 &    0.01 &   -3.12 & Star   \\
40 & CXOU043120.5+645122  & 04 31 20.51 & +64 51 22.5 &  178.5$\pm$13.9 &    10.5$\pm$3.6 &   36.5$\pm$6.5 &  132.5$\pm$12.1 & 0.94 &  $<$0.09 & \nodata &  $<$0.12 & \nodata &   22.23 &    0.16 &    2.31 & AGN    \\
41 & CXOU043121.1+645029  & 04 31 21.12 & +64 50 29.4 &    25.0$\pm$5.5 &     3.5$\pm$2.2 &   18.5$\pm$4.7 &     4.0$\pm$2.5 & 0.90 &  $<$0.09 & \nodata &  $<$0.12 & \nodata &   20.26 &    0.06 &   -1.45 & Star   \\
42 & CXOU043124.4+645115  & 04 31 24.47 & +64 51 15.5 &    72.6$\pm$9.2 &     4.1$\pm$2.5 &   18.1$\pm$4.7 &    51.3$\pm$7.8 & 0.87 &  $<$0.09 & \nodata &  $<$0.12 & \nodata & $>$23.7 & \nodata & $>$2.75 & AGN    \\
43 & CXOU043124.3+645121  & 04 31 24.38 & +64 51 21.2 &    18.8$\pm$5.7 &     1.8$\pm$0.9 &    2.0$\pm$1.8 &    21.0$\pm$5.5 & 0.78 &  $<$0.09 & \nodata &  $<$0.12 & \nodata &   21.26 &    0.07 &   -0.82 & Star   \\
44 & CXOU043125.1+645154  & 04 31 25.16 & +64 51 54.2 &  328.8$\pm$18.7 &     3.0$\pm$2.2 &    6.2$\pm$2.9 &  320.5$\pm$18.4 & 0.88 &  $<$0.09 & \nodata &  $<$0.12 & \nodata &   19.67 &    0.03 &    0.57 & AGN    \\
45 & CXOU043129.5+645053  & 04 31 29.52 & +64 50 53.3 &  130.2$\pm$12.1 &     4.0$\pm$2.8 &   17.2$\pm$4.6 &  109.8$\pm$11.1 & 1.29 &  $<$0.09 & \nodata &  $<$0.12 & \nodata &   20.91 &    0.05 &    0.78 & AGN    \\
\tablebreak
\enddata
\tablerefs{
(1) Reference ID \# for this paper;
(2) Chandra X-Ray Observatory identifier ; 
(3) J2000 Right Ascension from ACIS broadband image; 
(4) J2000 Declination; 
(5) Background-subtracted X-ray counts for each source in the 85 ksec observation 
	detected in 0.3-6 keV band and $1\sigma$ uncertainty from photon statistics; 
(6) Background-subtracted X-ray counts for each source in the 85 ksec observation 
	detected in broadband 0.3-0.7 keV soft X-ray band; 
(7) Background-subtracted X-ray counts for each source in the 85 ksec observation 
	detected in broadband 0.7-1.1 keV medium X-ray band; 
(8) Background-subtracted X-ray counts for each source in the 85 ksec observation 
	detected in broadband 1.1-6 keV hard X-ray band; 
(9) Ratio of source size to PSF size; 
(10) Flux of radio counterpart, if any, in mJy at 3.6 cm.  Nondetections are 
	listed as $3\sigma$ upper limits ; 
(11) $1\sigma$ uncertainty; 
(12) Flux of radio counterpart, if any, in mJy at 20 cm.  Nondetections are 
	listed as $3\sigma$ upper limits; 
(13) $1\sigma$ uncertainty; 
(14) R-band magnitude of optical counterpart, if any, from either our R-band imaging, or
	the HST F555W images from Hunter \et (2000);
(15) R-band $1\sigma$ uncertainty;
(16) Log of the ratio of X-ray to optical flux, $f_x/f_V$ : for this
purpose we use $f_x(erg~s^{-1} cm^{-2}) = 5.1\times10^{-18} x C_{tot}$ 
where $C_{tot}$ is the number of detected photons from column 5 and 
$f_V(erg~s^{-1} cm^{-2})=1.7\times10^{-9} \times 2.5^{-V}$
where $V$ is approximated by the R  magnitude from column 14.
(17) Probable source type based on X-ray, radio, optical fluxes or IDs. 
Objects with $log(f_x/f_V)<-0.5$ we tentatively identify as 
stars; Objects with $log(f_x/f_V)<-0.5$ we tentatively identify as 
probable AGN.  Objects in the disk of NGC~1569 without optical ID
we identify as probable X-ray binaries.  
}
\end{deluxetable}

%% file: table2.tex
\begin{deluxetable}{llrrrrrrr}  
\rotate
\label{pntsrcfits.tab}
\setlength{\tabcolsep}{0.02in} 
\tablewidth{6.0in}
\tabletypesize{\scriptsize} 
\tablecolumns{9} 
\tablecaption{Fitted Models for Two Brightest Compact X-ray Sources in NGC~1569}  
\tablehead{  
\colhead{Object}& \colhead{$\chi^2$}& \colhead{N(H)}                     & \colhead{$\Gamma$} & \colhead{kT}   & \colhead{$Z/Z_\odot$} & \colhead{$F_a(0.3-6)$}          & \colhead{$F_u(0.3-6)$}               & \colhead{$L_u(0.3-6)$} \\
\colhead{  }    & \colhead{}        & \colhead{$\times10^{22}~cm^{-2}$}  & \colhead{}         & \colhead{(keV)}& \colhead{}            & \colhead{erg cm$^{-2}$ s$^{-1}$} & \colhead{$erg~cm^{-2}~s^{-1}$}  & \colhead{$erg~s^{-1}$}  \\
\colhead{(1)}   & \colhead{(2)}     & \colhead{(3)}                      & \colhead{(4)}      & \colhead{(4)}  & \colhead{(6)}         & \colhead{(7)}                    & \colhead{(8)}                      & \colhead{(9)} }
\startdata  
\#16 Power Law       & 1.02 &  $0.80\pm0.19$            & $3.66\pm0.32$    & \nodata        & \nodata       &  5.7$\times10^{-14}$ & 47.7$\times10^{-14}$ & 3.1$\times10^{38}$\\
\#16 MEKAL           & 0.87 &  $0.15\pm0.14$            & \nodata          & $0.94\pm0.14$  & $0.03\pm0.01$ &  5.3$\times10^{-14}$ & 13.7$\times10^{-14}$ & 7.9$\times10^{37}$\\
\#16 Blackbody       & 1.46 &  $0.0\pm0.02$             & \nodata          & $0.28\pm0.02$  & \nodata       &  4.4$\times10^{-14}$ &  7.8$\times10^{-14}$ & 4.5$\times10^{37}$\\
\#19 Power Law       & 0.91 &  $0.72\pm0.32$            & $2.31\pm0.45$    & \nodata        & \nodata       &  5.0$\times10^{-14}$ &  9.5$\times10^{-14}$ & 5.4$\times10^{37}$\\
\#19 MEKAL           & 0.96 &  $0.12\pm0.08$            & \nodata          & $2.99\pm1.45$  & $0.34\pm0.41$ &  3.9$\times10^{-14}$ &  6.4$\times10^{-14}$ & 3.6$\times10^{37}$\\
\#19 Blackbody       & 1.41 &  $0.00\pm0.38$             & \nodata          & $0.42\pm0.05$  & \nodata       &  2.6$\times10^{-14}$ &  3.5$\times10^{-14}$ & 2.0$\times10^{37}$\\
\enddata  
\tablerefs{
(1) Model: All models include a forground Galactic absorbing component with solar metallicity and $N(HI)=2.0\times10^{21}~cm^{-2}$
and a foreground component with 0.25 $Z_\odot$ to represent the ISM within NGC~1569.;
(2) Reduced $\chi^2$ ; 
(3) best fit $Z=0.25~Z_\odot$ absorbing column density within NGC~1569 at 90\% confidence ; 
(4) best fit power law photon index at 90\% confidence ; 
(5) best fit temperature for MEKAL thermal plasma at 90\% confidence ; 
(6) best fit metallicity for MEKAL thermal plasma at 90\% confidence  ; 
(7) Absorbed Model flux from 0.3--6 keV ;
(8) Unabsorbed Model flux from 0.3--6 keV  ;
(9) Unabsorbed luminosity from 0.3--6 keV } 
\end{deluxetable}

%% file: fig_cap.tex
	\begin{figure}
	 \caption{X-ray point sources and star clusters in
the main stellar body of NGC~1569.  The greyscale shows a logarithmic
representation of the R-band optical image and the contours show the
adaptively-smoothed Chandra ACIS broadband 0.3-6 keV image.  Contours
levels are drawn every factor of 4 at 0.0005, 0.0020, 0.0080, 0.032, 0.128, 
0.5, 2.0, and 8.0 photons~s$^{-1}$ arcmin$^{-1}$.  
	}
        \label{fig:KNTR-R_XRAY-SM.PLOT} \end{figure}

	\begin{figure}
	 \caption{HST F555W broadband optical image (logarithmic greyscale)
with adaptively-smoothed X-ray image (contours) showing the central X-ray 
point sources and star clusters in the nucleus of NGC~1569. Contour levels 
are the same as Figure~\ref{fig:KNTR-R_XRAY-SM.PLOT}.  
	}
        \label{fig:KNTR-F555_XRAY.PLOT} \end{figure}

	\begin{figure}
	\caption{Two-color diagram for point sources.
	Lines represent the colors of model spectra.  
	(Upper panel) Sources projected against the main body of NGC 1569. 
	Power laws with photon indices of $\Gamma$=1.5, $\Gamma$=2.5, 
	$\Gamma$=3.5, and a blackbody spectrum with a temperature kT=0.2 keV 
	are shown for several foreground HI column densities.
	(Lower panel) Sources within a few arcminutes of the disk.
	MEKAL thermal plamsas with temperatures kT=1.0 and kT=2.0 keV are 
	shown for several foreground HI column densities.
  	Objects identified as probable stars in Table~1 on the basis
	of $f_X/f_V$ are indicated.
	}
  	\label{fig:2color} \end{figure}

	\begin{figure}
	\caption{Histogram of the hard-band (1.1--6 keV) source fluxes 
for 14 point sources in \n1569.  The solid 
line shows a cumulative luminosity function with slope $\alpha=0.5$ 
normalized to the highest luminosity bin (not a fit).  
	}  \label{fig:hardl_hist}  \end{figure}

	\begin{figure}
	 \caption{Contours of 0.3-6~keV \x emission on \Ha image.  
The contours are spaced by a factor of 1.347 from 0.00594 to 1.696 
cnts~s~$^{-1}$ per square arcminute.  The \Ha image is displayed on 
logarithmic scale, and 
the filaments identified by Hunter \et (1993) are marked. 
White bar represents 1\farcm0.
	}
        \label{fig:all2_ha} \end{figure}

\clearpage

	\begin{figure}
	 \caption{X-ray contours on \Ha. 
Letters mark regions of maximal expansion velocity in the \Ha nebula
from Martin (1998).  Note that the fingers of \x emission protrude into
these shells.  Apertures discussed in the paper are marked in
green.  Circle marks the position of SSC~A.  The length of the rectangular
apertures is 1~kpc.
	}
        \label{fig:all2_ha_inset} \end{figure}

   	\begin{figure}
         \caption{Composite 3-color image of NGC~1569 with the Chandra 
0.3-6~keV \x emission in green, \Ha emission in red, and optical 6450\AA\ 
continuum in blue.    Contours show the 21-cm neutral hydrogen
column density at levels of $1\times10^{21}~cm^{-2}$ (bold),
$4\times10^{21}~cm^{-2}$ (solid line), and $7\times10^{21}~cm^{-2}$
(dashed line).
        }  
        \label{fig:haxr} \end{figure}

   	\begin{figure}
         \caption{Chandra true-color \x image of \n1569.  
Red, green, and blue represent the soft (0.3 - 0.7 keV), medium (0.7 - 
1.1 keV), and hard (1.1 - 6 keV) regions of the Chandra bandpass.
The images have been adaptively smoothed (see text). Contours
are the same as in Figure~\ref{fig:haxr}.
        }  
        \label{fig:3x} \end{figure}

   	\begin{figure}
         \caption{Contours of soft color $C_1=(M-S)/(M+S)$ on 
broadband \x intensity.  The softest regions, $C_1 \approx 0.3~{\rm to}~
0.5$, correspond to the faintest regions of the halo.
        }  
        \label{fig:mscon_x} \end{figure}

   	\begin{figure}
         \caption{Integrated spectrum of \n1569.  The contribution
of resolved point sources is shown in blue.  The red line shows the
spectrum of the diffuse component.  Prominent lines of O, Ne, Mg, and Si
are marked.  The Fe~L lines are largely blended, so we have marked 
only the energies of the strongest line from each ionization state
FeXVII through FeXXIV.
The blue line represents the background simulated from a set of deep
exposures (see \S~2.1.1).  
The OVII and OVIII lines are prominent 
whichever mode of background-subtraction is employed, but some of the
features near Si lines may be residuals of the background subtraction.
        }  
        \label{fig:int} \end{figure}

   	\begin{figure}
         \caption{Integrated spectrum of \n1569.  The folded spectrum
	and residuals are shown for several models from Table
	3.  See discussion in the text.
        }  
        \label{fig:ellip_fit} \end{figure}

\clearpage

   	\begin{figure}
         \caption{
	Count distributions of the {\it red} and {\it green} regions
	of the southern halo (cf. \fig~\ref{fig:3x}). The physical
	parameters of the emitting plasma are dicussed in \S~3.2.2
	and \S~3.3.3.
        } \label{fig:rg_spec} \end{figure}

	\begin{figure}
	\caption{
	Geometry of the disk and outflow. 
	} \label{fig:geometry} \end{figure}

   	\begin{figure}
         \caption{Comparison of the diffuse emission spectrum on the 
	south and north side of the disk.  Spectra from different
	apertures are offset for ease of comparision, and the northern 
	spectra have been scaled to match the 1 to 2~keV counts in the 
	southern spectra. 
        }  
        \label{fig:ns} \end{figure}

   	\begin{figure}
         \caption{Equivalent hydrogen absorption column, fitted to
	the thermal emission model, versus projected distance from the 
	plane of the disk. The total column includes absorption from the 
	disk of \n1569 and foreground Galactic absorption.  The results
	are shown for three hypothetical values of the gas-phase metallicity in \n1569.
	The foreground column is assumed to have solar metallicity.
	The thick, solid line represents the total Galactic and 
	\n1569 hydrogen columns measured from 21-cm emission. 
	The error bars represent the 90\% confidence range for the
	assumed two component model.   We assume solar abundances
        (Anders \& Ebihara 1982) 
        for the foreground, Galactic gas but vary the 
        metallicity of the intrinsic absorbing column of the \n1569 disk.
        }  
        \label{fig:zH} \end{figure}

	\begin{figure}
	\caption{The tradeoff between metallicity and normalization
	of the thermal components in our models of the integrated
	spectrum. The blue, green, and red lines illustrate
	the hard thermal, soft thermal, and power law components for
	two  different models from Table~5 --
	the $1.0 Z_{\alpha,\odot}$ model in panel~a and the
	$100 Z_{\alpha,\odot}$ model in panel~b.
	The spectra resulting from these two
	models are difficult to distinguish.
	}
	\label{fig:ac} \end{figure}
	
	\begin{figure}
	\caption{Fits to the integrated spectrum (points) with 
	variable abundances.  Folded spectra and residuals 
	illustrate three models from Table~5:  $Z_{\alpha} = 
	100~Z_{\alpha,\odot}$ (black), 1.0~$Z_{\alpha,\odot}$ (blue), 
	and 0.1~$Z_{\alpha,\odot}$ (magenta).  The spectral normalizations
	vary a great among these models as illustrated in Table~5.
	}
	\label{fig:alpfe} \end{figure}

\clearpage

	\begin{figure}
	\caption{Relative contribution of alpha elements and iron
	to the spectral model {\it wabs(vabs*Pt + vabs*M + vabs*M)}.
	The $Z_{\alpha} = 1.0 Z_{\alpha,\odot}$ from
	Table~5 is compared to models with little iron (0.1$Z_{Fe,\odot}$) 
	(red) and a model with a paucity of alpha elements 
	(0.04$Z_{\alpha,\odot}$) 
	(green). (a) Prominent lines are labeled for the models. (b)
	Changing the ratio of alpha elements to iron has a measurable effect
	on the folded spectrum.  
	}
	\label{fig:amod2_adat2} \end{figure}

	\begin{figure}
	\caption{Elemental abundances fitted to the spectrum
	of the diffuse gas. The cross marks the best fit. Contours 
	represent the 60\%, 90\%, and 99\% 
	confidence levels. The O/H and Fe/H abundance ratios are shown
	relative to the solar photospheric values of Anders \& Grevesse (1989).
	On the more common solar meteoritic scale (Anders \& Grevesse 
	1989), the relative Fe abundances are 0.085 dex higher.
	Note the Mg, Ne, Si, and Ca abundances are tied to O, and
	the other metals vary with the Fe abundance. 
	}
	\label{fig:con93} \end{figure}

   	\begin{figure}
         \caption{Implications of the spectral degeneracy between the 
	fitted emission	measure and metallicity of the diffuse hot gas.
	(Top) Dependence of the inferred gas mass on the assumed alpha
	element abundance (in solar units). (Bottom) The inferred mass 
	of oxygen in the wind increases slowly as the alpha element
	abundance of the spectral model is increased, i.e.  
	$M_O \propto Z_{\alpha}^{0.55}$.
	 } \label{fig:mo} \end{figure}

   	\begin{figure}
         \caption{Entraining interstellar gas in the
	   hot supernova-driven wind affects the O abundance and the
	   O/Fe abundance ratio in the wind.  As the mass loading
	   increases (toward the right), the O and O/Fe decrease
	   toward their values in the ISM (see \S~5.2.1). Three
	   tracks are shown to illustrate the uncertainty in the
	   supernova yields.
	Using the relation between $M_x$ -- $Z_{\alpha}$ from the fitted
	spectral models, the curves can be parameterized by the implied
	mass of stellar ejecta at each point.  At a given metallicity
	(and therefore wind mass), a high mass loading factor implies
	a low mass of stellar ejecta in the wind.  The solid circles
	delimit the mass of stellar ejecta produced by the starburst 
	model after a time of 10~Myr ($9.8 \times 10^4$\msun) and 
	20~Myr ($3.8 \times 10^5$\msun). 
	The Chandra measurements for O and O/Fe combined
	with the Starburst~99 estimate for the mass of supernova ejecta
	defines the data points and error bars.
	Consistent models require substantial mass loading of $\chi \sim 9$.
} \label{fig:chem} \end{figure}

   	\begin{figure}
         \caption{
Model chemical enrichment history as a function of gas mass fraction.
The solid line shows a closed box model assuming instantaneous recycling.  
The dashed lines show the evolutionary tracks if 75\%, 50\%, and 25\% of the 
oxygen or an equivalent fraction of the gas mass is lost from the galaxy.  The
large symbol with error bars shows the {\it full permitted range}
of observed gas mass fraction ($0.35 < \mu < 0.73$; see text)
and ISM metallicity of NGC~1569 from HII regions ($0.2< Z/Z_\odot < 0.25$).  
       } \label{fig:chemev3} \end{figure}